\documentclass[aps,prd,reprint,superscriptaddress,nofootinbib, longbibliography]{revtex4-1}

\usepackage{amsmath,amssymb, amsthm,amstext}
\usepackage{natbib}
\usepackage{graphicx}
\usepackage{color}
\usepackage{array, enumerate}
\usepackage{bm}
\usepackage{multirow}
\usepackage[breaklinks,colorlinks,citecolor=blue]{hyperref}
\usepackage{braket}
\usepackage{txfonts}
\usepackage[normalem]{ulem}

\def\be{\begin{equation}}
\def\ee{\end{equation}}

\def\apjl{Astroph.J.Lett.}
\def\apjs{Astroph.J.S.}
\def\mnras{MNRAS}

\def\aap{Astronomy\&Astrophysics}

\begin{document}

\title{Probing orbits of stellar mass objects deep in galactic nuclei with quasi-periodic eruptions}
\author{Cong Zhou}
\email{dysania@mail.ustc.edu.cn}
\affiliation{CAS Key Laboratory for Research in Galaxies and Cosmology, Department of Astronomy, University of Science and Technology of China, Hefei 230026, P. R. China}
\affiliation{School of Astronomy and Space Sciences, University of Science and Technology of China, Hefei 230026, P. R. China}
\author{Lei Huang}
\affiliation{Shanghai Astronomical Observatory, Chinese Academy of Sciences, 80 Nandan Road, Shanghai 200030, People’s Republic of China}
\affiliation{Key Laboratory for Research in Galaxies and Cosmology, Chinese Academy of Sciences, Shanghai 200030, People’s Republic of China}
\author{Kangrou Guo}
\affiliation{Tsung-Dao Lee Institute, Shanghai Jiao-Tong University, Shanghai, 520 Shengrong Road, 201210, People’s Republic of China}
\affiliation{School of Physics \& Astronomy, Shanghai Jiao-Tong University, Shanghai, 800 Dongchuan Road, 200240, People’s Republic of China}
\author{Ya-Ping Li}
\affiliation{Shanghai Astronomical Observatory, Chinese Academy of Sciences, 80 Nandan Road, Shanghai 200030, People’s Republic of China}
\affiliation{Key Laboratory for Research in Galaxies and Cosmology, Chinese Academy of Sciences, Shanghai 200030, People’s Republic of China}
\author{Zhen Pan}
\email{zhpan@sjtu.edu.cn}
\affiliation{Tsung-Dao Lee Institute, Shanghai Jiao-Tong University, Shanghai, 520 Shengrong Road, 201210, People’s Republic of China}
\affiliation{School of Physics \& Astronomy, Shanghai Jiao-Tong University, Shanghai, 800 Dongchuan Road, 200240, People’s Republic of China}
\begin{abstract}
 Quasi-periodic eruptions (QPEs) are intense repeating soft X-ray bursts with recurrence times about a few to ten hours from nearby galactic nuclei. The origin of QPEs is still unclear.  In this work, we investigated the  extreme mass ratio inspiral (EMRI) + accretion disk model, where the disk is formed from a previous tidal disruption event (TDE). In this EMRI+TDE disk model, the QPEs are the result of collisions between a TDE disk and a stellar mass object (a stellar mass black hole or a main sequence star) orbiting around a supermassive black hole (SMBH) in galactic nuclei. 
 If this interpretation is correct, QPEs will be invaluable in probing the orbits of stellar mass objects in the vicinity of SMBHs, and further inferring the formation of EMRIs which are one of the primary targets of spaceborne gravitational wave missions. Taking GSN 069 as an example, we find the EMRI wherein is of low eccentricity ($e<0.1$ at 3-$\sigma$ confidence level) and semi-major axis about $O(10^2)$ gravitational radii of the central SMBH, which is consistent with the prediction of the wet EMRI formation channel,
 while incompatible with alternatives.
\end{abstract}
\date{\today}

\maketitle

\section{Introduction}

In the past decade, X-ray quasi-periodic eruptions (QPEs) have been detected in nearby galactic nuclei 
\cite{Sun2013,Giustini2020,Arcodia2021,Arcodia2022,Chakraborty2021} which host low-mass 
($\simeq 10^5-10^7 M_\odot$ at most) central supermassive black holes (SMBHs) \cite{Wevers2022,Miniutti2023}. QPEs are 
fast bright soft X-ray bursts repeating every few hours with peak X-ray luminosity $10^{42}-10^{43}$ ergs s$^{-1}$.
QPEs have  thermal-like X-ray spectra with temperatures $kT\simeq 100-250$ eV, in contrast with the  
temperatures $\simeq 50-80$ eV in the quiescent state.  The presence of a narrow line region in all QPE host galaxies 
implies that a long-lived active galactic nucleus (AGN) likely plays an integral role in the QPEs, while
the absence of luminous broad emission lines indicates that none of the central SMBHs are currently actively accreting, 
i.e., they are all likely recently switched-off AGNs
\cite{Wevers2022}. In addition, QPEs similar to tidal disruption
events (TDEs) are preferentially found in post-starburst galaxies \cite{Wevers2022},
and two QPE sources (GSN 069 and XMMSL1 J024916.6-04124) and a candidate (AT 2019vcb), have been directly associated with
X-ray TDEs \cite{Shu2018,Sheng2021,Chakraborty2021,Miniutti2023,Quintin2023}.
A recent XMM-Newton observation of GSN 069 identified the reappearance of QPEs after two years of absence \cite{Miniutti2023}.
This observation result shows that QPEs may only be present below a quiescent luminosity threshold $L_{\rm thr}\sim 0.4 L_{\rm Edd}$, where $L_{\rm Edd}$ is the Eddington luminosity, and a new phase shows up  in the QPE reappearance
where the intensity and the temperature of two QPEs become different.
Long term observation of yields more interesting features of GSN 069 \cite{Miniutti2023b}, e.g., the quasi-periodic oscillations (QPOs) in 
the quiescent state following the QPEs, long-term evolution of the quiescent level emission is consistent with a TDE 
or even possibly a repeating TDE, 
and QPEs measured in higher energy bands are stronger, peak earlier, and have shorter
duration than when measured at lower energies.

Many models have been proposed for explaining the physical origin of QPEs, 
based on different disk instabilities\footnote{The classical limit-cycle instability in AGN disks is disfavored because the the QPE periods, burst timescales and burst profiles cannot be reconciled with the limit-cycle prediction  \cite{Arcodia2021,Liu2021}. 
}  \cite{Raj2021,PanX2022,Panx2023,Kaur2023,Snieg2023}, 
self-lensing binary massive black hole \cite{Ingram2021},
mass transfer at pericenter from stars or white dwarfs orbiting around the SMBH
\cite{King2020,King2022,King2023,Chen2022,Wang2022,Zhao2022,Metzger2022,Lu2022,Krolik2022,Linial2023b},
(periodic) impacts between a stellar-mass object (SMO), a star or a stellar mass black hole (sBH),  and the accretion disk that is formed following
a recent TDE (dubbed as EMRI+TDE disk model)  \cite{Sukova2021,Xian2021,Tagawa2023, Linial2023, Franchini2023}.
These models can explain some of the features in the QPE light curves (mainly GSN 069),  
there is no model producing diverse features of different QPE sources (see more discussions in e.g.  \cite{Miniutti2019,Giustini2020,Arcodia2021,Arcodia2022,Chakraborty2021,Wevers2022,Miniutti2023,Miniutti2023b}). 

Recently, Linial et al. \cite{Linial2023} and Franchini et al.  \cite{Franchini2023} pointed out the EMRI + TDE disk model 
is flexible  in recovering comprehensive features of different QPEs. 
In this EMRI+TDE disk model, the stellar mass object (SMO) could be a sBH  or a main sequence star, and the two share many similar model predictions, e.g., the long-short pattern in the QPE recurrence times and 
 the strong-weak pattern in the QPE intensities (see \cite{Franchini2023,Linial2023,Linial2023d} for details),
 therefore it seems hard to distinguish the two with the existing QPE observations.
In the majority of this work, we tend not to distinguish them and we will briefly discuss their different predictions that might be tested by near future observations in the final part of this paper.
We first refine the flare model following the analytic supernova explosion model developed by Arnett \cite{Arnett1980}, 
and have been used in modelling the optical flares of OJ 287 \cite{Lehto1996,Pihajoki2016} and SDSSJ1430+2303 \cite{Jiang2022}.
With this refinement, we find the model fitting to the QPE light curves with reasonable precision though the effective temperatures at the light curve peaks in the best-fit model is in mild tension with the observation values.
Due to this limitation of the plasma ball emission model, we consider an alternative phenomenological model
where the light curve consists of a rising part and a decay part with different timescales.
We apply both models to the QPEs from GSN 069 and find out the starting time of each flare,
with which we constrain the EMRI orbital parameters.
We find the EMRI orbit inferred from these QPEs is of low eccentricity ($e<0.1$ at 3-$\sigma$ confidence level) 
and semi-major axis $a= \mathcal{O}(10^2)M_\bullet$,
where $M_\bullet$ is the gravitational radius of the central SMBH. If the EMRI+TDE disk is indeed the origin of QPEs, the
EMRI orbital parameters inferred from the QPEs will be invaluable in probing the EMRI formation channels. 
As for the GSN 069 EMRI, we find it is highly unlikely comes from the loss-cone channel or the Hills mechanism,
while is consistent with the wet channel expectation (see Figs.~\ref{fig:no_lc}, ~\ref{fig:sBH_AGN}).

This paper is organized as follows. In Section~\ref{sec:model} , we briefly review the flare emission mechanism,
introduce two analytic model for fitting the QPE light curves, and the EMRI equations of motion (EoM). 
In Section~\ref{sec:application} , we fit the QPE light curves finding out the starting time of each flare,
with which we constrain the EMRI orbital parameters.  
In Section~\ref{sec:summary}, we summarize this paper by evaluating the performance of the EMRI+disk model predictions, discussing the implications of QPEs on EMRI formation channels, possible observable signatures for distinguishing stellar 
EMRIs versus sBH EMRIs and future work.
In Appendix~\ref{app:corner}, we include the detailed corner plots of emission model parameters and the flare timing model parameters.
In Appendix~\ref{app:scatter}, we analyze the orbital stability of the SMO after a possible close-by scattering with a TDE remnant star.
In Appendix~\ref{app:hd_sim}, we show some hydrodynamic (HD) simulation results of SMO-disk collisions and infer possible observational features 
in the resulted light curves.
In this paper, we use geometrical units with $G=c=1$ if not specified otherwise.

\section{QPE model: EMRI+TDE disk}\label{sec:model}

\begin{figure*}
\includegraphics[scale=0.48]{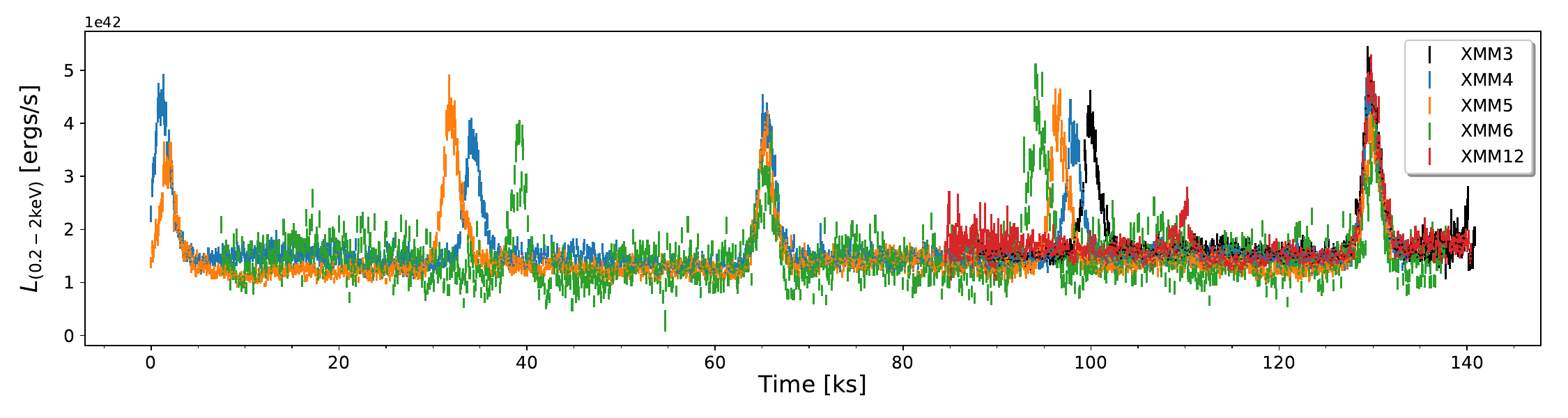}
\caption{\label{fig:raw_lc} The QPE light curves from observations XMM 3-6 and 12, where we have aligned different observations at their last peak locations, and we have also rescaled the XMM 6 luminosity.}
\end{figure*}

Fig.~\ref{fig:raw_lc} displays the GSN 069 QPEs found in 5 XMM Newton observations.
The peak luminosity, recurrence time, and duration of QPEs are $\mathcal{O}(10^{42})$ ergs/s,
$\mathcal{O}(10)$ hours and $\mathcal{O}(1)$ hour, respectively. 
In this section, we will first sketch the EMRI+TDE disk model identifying the parameter space that 
is consistent with above energy budget and time scales, then explain the details of the emission models 
for fitting each individual QPE light curve and the EMRI model for fitting the starting times of all QPEs.

\subsection{Emission model}
As a the SMO orbiting around the SMBH and cross the accretion disk, 
the relative velocity between the SMO and the local accretion flow 
in general is higher than the local sound speed $v_{\rm vel} > c_{\rm s}$,
and the gas in the disk will be compressed and heated by the shock wave.
For sBHs, the accretion radius $r_{\rm acc}:=2Gm/v_{\rm rel}^2$ is in general much larger than the geometrical size $m$. 
As shown by simulations in Ref.~\cite{Ivanov1998}, the heated gas expands along the shocked tunnel 
and forms a hot, optically thick, radiation-dominated plasma ball on each side of the disk.
The hot plasma ball cools down due to (nearly) adiabatic expansion and thermal radiation. 
For stars, the whole process is similar except now the geometrical radius $R_\star$ plays the role of the accretion radius in 
the sBH case. 

We consider the standard $\alpha$ disk model \cite{SS1973}, 
where the disk structure in the radiation dominated regime can be analytically expressed as \cite{Kocsis2011}
\be 
\begin{aligned}
    \Sigma(r) 
    &= 1.7\times 10^5 \ {\rm g\ cm}^{-2} \alpha_{0.01}^{-1} \dot M_{\bullet,0.1}^{-1} r_{100}^{3/2}\ , \\ 
    H(r) &= 1.5 M_\bullet  \dot M_{\bullet,0.1}\ ,
\end{aligned}
\ee 
where  $H$ is the scale height of the disk from the mid-plane, $\Sigma$ is the disk surface density
and we have defined $\alpha_{0.01}=\alpha/0.01, r_{100}=r/100 M_\bullet, \dot M_{\bullet,0.1}=\dot M_\bullet/(0.1 \dot M_{\bullet,\rm Edd})$ 
with $\dot M_{\bullet,\rm Edd}$ the Eddington accretion rate.
The disk surface density and the disk thickness will be used in the energy budget estimates 
and the EMRI orbit inference.

For a star, the energy loss after crossing a disk is simply 
\be \label{eq:delta_E_star}
\begin{aligned}
     \delta E_{\star} 
     &= 2\times \frac{1}{2} \delta m_{\rm gas} v_{\rm rel}^2 \, \\
     &\approx 3\times 10^{46} {\rm ergs}\times \Sigma_5
     R_{\star,\odot}^2   r_{100}^{-1} \sin\iota_{\rm sd}\ ,
\end{aligned}
\ee 
where the factor $2$ in the first line takes both the thermal energy and the kinetic energy of the shocked gas into account, 
$\Sigma_5 :=\Sigma/(10^5 {\rm g \ cm^{-2}}), R_{\star,\odot}:=R_\star/R_\odot$, $\delta m_{\rm gas} = 2\pi R_\star^2\Sigma/\sin\iota_{\rm sd}$ is the mass of the shocked gas \cite{Linial2023},
and we have used the approximation
$v_{\rm vel} \approx v_{\rm K}\sin\iota_{\rm sd}$ with $v_{\rm K}$ the local Keplerian velocity,
where $\iota_{\rm sd}$ is the angle between 
the sBH orbital plane and the disk plane.

For a sBH moving in a uniform gas cloud, the gravitational drag from the perturbed gas is \cite{Chand1943,Rephaeli1980,Binney1987,Ostriker1999}
\be\label{eq:drag}
\begin{aligned}
    \mathbf{F}_{\rm drag}
    &= -4\pi\ln\Lambda \frac{G^2\rho m^2}{v^3_{\rm rel}}
    \left[{\rm erf}(X)-\frac{2X}{\sqrt{\pi}}e^{-X^2} \right] \mathbf{v}_{\rm rel}\ , \\
    &\approx -4\pi\ln\Lambda \frac{ G^2\rho m^2}{v^3_{\rm rel}}
    \mathbf{v}_{\rm rel}\ ,
\end{aligned}
\ee 
where $\rho$ is the gas density; $m$ is the sBH mass;
$X=v_{\rm rel}/(\sqrt{2} c_s)$ is the ratio of relative velocity $v_{\rm rel}$ over the gas particle velocity dispersion (approximated by the local sound speed $c_s$), and we have used the approximation $v_{\rm rel}/c_s \gg 1$ in the second line;
 $\ln\Lambda= \ln(b_{\rm max}/b_{\rm min})$ is the Coulomb logarithm
with $b_{\rm max/min}$ the maximum/minimum cutoff distance associated to the interaction.
The cutoff distance $b_{\rm max}$ is the maximum extent of the wake $2H\sin^{-1}\iota_{\rm sd}$, 
while $b_{\rm min}$ was identified either as the 
 accretor size $m$ \cite{Shankar1993, Ruffert1994, Edgar2004,Chapon2013} or as the stand off distance $r_{\rm so}\approx \frac{1}{2}r_{\rm acc}$ \cite{Thun2016}.
With these two different identifications, we have $\ln\Lambda   \approx 13 \ {\rm or}\  5$, 
where we have used the fiducial 
values $M_\bullet=10^6 M_\odot, m=30 M_\odot, H=1.5 M_\bullet$ and $\sin\iota_{\rm sd}=0.1$  in the approximate equal sign.
As a result, the sBH loses energy after crossing a disk \footnote{In estimating the sBH energy loss in crossing the disk, people easily use the approximation $\delta E_{\rm sBH} = \frac{1}{2} \delta m_{\rm gas} v_{\rm rel}^2$. In fact,
$\delta m_{\rm gas}$ above is merely the amount of gas that is supposed to be accreted by the sBH, which is only
a fraction of shocked gas (the shock size is larger than the accretion radius $r_{\rm acc}$ by quite a few times \cite[see e.g.][]{Hunt1971,Hunt1979,Shima1985,Ruffert1994,Ruffert1996}). The more accurate estimate of the sBH energy loss would be  Eq.~(\ref{eq:delta_E}), which was derived in Ref.~\cite{Rephaeli1980} and had been verified with hydrodynamic simulations \cite[see e.g.][]{Shima1985,Shankar1993,Ruffert1994,Ruffert1996,Chapon2013}. }
\be \label{eq:delta_E}
\begin{aligned}
    \delta E_{\rm sBH} 
    &= F_{\rm drag} \Delta L   
    = 4\pi \ln\Lambda  \frac{G^2m^2}{v_{\rm rel}^2}\frac{\Sigma}{\sin(\iota_{\rm sd})}\ ,\\ 
    &\approx 2\times10^{46} {\rm ergs} \left(\frac{\ln\Lambda}{10}\right) \Sigma_5 m_{30}^2 r_{100} \left(\frac{\sin\iota_{\rm sd}}{0.1}\right)^{-3} \ ,
\end{aligned}
\ee 
where 
$\Delta L = 2H/\sin(\iota_{\rm sd})$ is the length of the sBH orbit inside the disk,
and we have defined $m_{30} = m/30 M_\odot$.

In order to find out the starting time of each flare,  
we consider the following two emission models: an expanding plasma ball emission model and a 
phenomenological model.

\subsubsection{Plasma ball model}
In general, the post-shock gas is not uniformly compressed, heated or accelerated depending on where it crosses the shock front \cite[see e.g.][for detailed simulations]{Hunt1971,Hunt1979,Shima1985,Ruffert1994,Ruffert1996}. For modelling the emission of the shocked gas, we simplify 
the shocked gas as a uniform plasma ball with initial size $R_{0}$ and initial surface temperature $T_{\rm e0}$.
Following Arnett  \cite{Arnett1980}, We model the plasma ball expansion, cooling and radiation 
as follows.

Considering a spherical plasma ball expanding uniformly, the evolution of the expanding shell follows the first law of thermodynamics, which states, in the Lagrangian coordinates,
\be \label{eq:first_law}
\dot{E}+P\dot{V}=-\frac{\partial L}{\partial m}
\ee
where $E$, $V$ are the specific energy and volume, $P$ and $L$ are the pressure and the radiation luminosity,
and $m$ is the total mass enclosed by the shell considered. In the diffusion approximation, $L$ is
\be \label{eq:L_T_gradiant}
\frac{L}{4\pi r^2}=-\frac{ac}{3\kappa\rho}\frac{\partial T^4}{\partial r}
\ee
where $\kappa$ is the  opacity dominated by Thomson scattering. 

Adiabatic homologous expansion of a photon-dominated gas ($\gamma=4/3$) gives $T\propto R(t)^{-1}$ and $\rho(t)\propto R(t)^{-3}$, where $R(t)$ is the boundary of the plasma ball. With these two relations, we can now write
\begin{align}
    \label{eq:T_seperation}
    T(x,t)^4&=\Psi(x)\phi(t)T_{00}^4\frac{R_0^4}{R(t)^4}\\
    \label{eq:rho_separation}
    \rho(t)&=\rho_0\frac{R_0^3}{R(t)^3}\\
    \label{eq:VdotoV}
    \frac{\dot{V}}{V}&=3\frac{\dot{R}}{R}
\end{align}
where $x=r/R(t)$ is a dimensionless comoving radial coordinate. In the time scale of our consideration, the outer boundary of the plasma ball is expanding at a constant speed
\be \label{eq:R_t}
R(t)\equiv R_0\left(1+\frac{t}{\tau_h}\right)=R_0+v_{sc}t
\ee
$\tau_h$ is the expansion timescale and $v_{sc}=R_0/\tau_h$ is the expanding velocity scale. In \eqref{eq:T_seperation}, we separate the space and time dependence of temperature. 

Plugging \eqref{eq:L_T_gradiant} to \eqref{eq:VdotoV} into \eqref{eq:first_law} and separate the PDE, we have
\begin{align}
    \label{eq:Psi_eq}
    \alpha&=-\frac{1}{x^2\Psi}\frac{\mathrm{d}}{\mathrm{d}x}\left(x^2\frac{\mathrm{d}\Psi}{\mathrm{d}x}\right)\\
    \label{eq:phi_eq}
    \frac{\mathrm{d}\phi}{\mathrm{d}t}&=-\frac{R(t)\phi}{R_0\tau_0}
\end{align}
where $\alpha$ is the eignvalue determined by the boundary condition, $\tau_0\equiv\frac{3\rho_0R_0^2\kappa}{\alpha c}$ is the diffusion time scale. We consider the 'radiative zero' boundary condition \cite{Arnett1980} ($\Psi(1)=0$). Together with the trivial boundary condition at the center $\Psi(0)=1,\,\mathrm{d}\Psi/\mathrm{d}x=0$, the solution of \eqref{eq:Psi_eq} is
\be \label{eq:sol_Psi}
\Psi(x)=\frac{\sin(\pi x)}{\pi x}
\ee
the eigenvalue $\alpha=\pi^2$. The time evolution equation \eqref{eq:phi_eq} can be solved with the trivial initial condition $\phi(0)=1$, whose solution is
\be \label{eq:sol_phi}
\phi(t)=\exp{\left[-\frac{t}{\tau_0}\left(1+\frac{t}{2\tau_h}\right)\right]}
\ee

With \eqref{eq:R_t}, \eqref{eq:sol_Psi} and \eqref{eq:sol_phi}, the temperature distribution and evolution is obtained. We now return to the surface luminosity. Plugging \eqref{eq:T_seperation}, \eqref{eq:R_t}, \eqref{eq:sol_Psi}, \eqref{eq:sol_phi} into \eqref{eq:L_T_gradiant} and take $x=1$, we have
\be \label{eq:surface_L}
L(1,t)=\frac{4\pi R_0 aT_{00}^4c}{3\kappa\rho_0}\phi(t)\equiv L(1,0)\phi(t)
\ee
Afterwards, we choose the effective temperature as the parameter of our model, which can be directly obtained from the surface luminosity
\be \label{eq:T_eff}
    T_{\rm e}=\left(\frac{L(1,0)}{4\pi\sigma R_0^2}\right)^{1/4}\phi^{1/4}\Bigg/\sqrt{1+\frac{t}{\tau_h}}\equiv T_{\rm e0}\phi^{1/4}\Bigg/\sqrt{1+\frac{t}{\tau_h}}
\ee

Applying this model to QPEs, we find the soft X-ray (0.2-2keV) luminosity 
\be \label{eq:model}
L_X=\frac{8\pi^2 hR^2}{c^2}\int_{0.2\mathrm{keV}}^{2\mathrm{keV}}\frac{\nu^3d\nu}{e^{h\nu/k_BT_\mathrm{e}}-1}\ ,
\ee
where the effective temperature $T_{\rm e}$ is determined by \eqref{eq:T_eff} and the plasma ball radius $R$ is determined by \eqref{eq:R_t}. The intrinsic parameters of our model are $\{R_0,\, T_{\rm e0},\,\tau_0,\,\tau_h\}$ in addition to the flare staring time $t_0$. For the purpose of model parameter inference, we find parameters $\{t_0, R_0,\, T_{\rm e0},\,\tau_m:=\sqrt{\tau_0 \tau_h}, \,v_{sc}:=R_0/\tau_h\}$ are better constrained, which are chosen as the model intrinsic parameters hereafter.

\subsubsection{Phenomenological model}
In additional to the above expanding plasma ball emission model, we also consider the following alternative phenomenological light curve model \cite{Norris2005, Arcodia2022}
\be \label{eq:phen}
L_{\rm X}(t) =
    \begin{cases}
     0  & \text{if $t\leq t_p-t_{\rm as}$}\\
      L_p e^{\sqrt{\tau_1/\tau_2}} e^{\tau_1/(t_p-t_{\rm as}-t)} & \text{if $t_p-t_{\rm as} < t<t_p$}\\
      L_p e^{-(t-t_p)/\tau_2} & \text{if $t\geq t_p$}
    \end{cases}       
\ee 
where $t_{\rm as}=\sqrt{\tau_1\tau_2}$. 
Following Ref.~\cite{Norris2005}, we define the flare  starting time as when 
 the flux is $1/e^3$ of the peak value, i.e.,  $L_{\rm X}(t_0) = L_p/e^3$.
Therefore only 3 out of the 4 time variables are independent, and we take $\{t_0, t_p, \tau_2\}$ and $L_p$ 
as the independent model parameters.

In addition to the QPEs, quasi-periodic oscillations (QPOs) has been identified in the quiescent state luminosity, we therefore model the background 
luminosity as 
\be  \label{eq:bgd}
L_{\rm bgd} (t)=B+A \sin\left(2\pi (t-t_0)/P_{\rm QPO} + \phi_{\rm QPO}\right)\ ,
\ee 
where $B$ is average background luminosity, $A, P_{\rm QPO}, \phi_{\rm QPO}$ are the QPO amplitude, period and initial phase, respectively.

\subsection{Flare timing}
In general, the sBH collides the accretion twice per orbit, and the propagation times of two flares produced by the collisions
to the observer are different due to different propagation paths. 

For convenience, we model the motion of the sBH as a geodesic in the Schwarzschild spacetime.
Considering an orbit with semimajor axis $a$ and eccentricity $e$, 
the pericenter and apocenter distance $r_{\rm a,p}=a(1\pm e)$ are the roots to 
the effective potential \cite{Chandrasekhar1983}
\be 
V(r) = r^4 E^2-(r^2-2r)(r^2+L^2)\ ,
\ee 
where all the radii/distances are formulated in units of the graviational radius $M_\bullet$.
From the effective potential, it is straightforward to obtain the orbital energy and angular momentum $E(a, e), L(a, e)$.
The equations of motion (EoMs) in the Schwarzschild spacetime can be derived from the Hamiltonian 
\be 
\mathcal{H}(r, \theta, p_r, p_\theta) =\frac{1}{2}g^{\mu\nu}(r, \theta) p_\mu p_\nu \ ,
\ee 
where $g^{\mu\nu}$ is the Schwarzschild metric.
Considering the simple case where the orbit lies on the equator, we obtain 
\be 
\begin{aligned}
    \dot r &= \left(1-\frac{2}{r} \right) p_r \left(\frac{dt}{d\tau}\right)^{-1}\ , \\
    \dot \psi &= \frac{L}{r^2} \left(\frac{dt}{d\tau}\right)^{-1} \ , \\ 
    \dot p_r &= -\left( \frac{p_r^2}{r^2} - \frac{L^2}{r^3} + \frac{E^2}{(r-2)^2} \right) \left(\frac{dt}{d\tau}\right)^{-1} \ ,
\end{aligned}
\ee 
where $dt/d\tau = E/(1-2/r)$ and  dots are the derivative w.r.t. to time $t$ and $\psi$ is the azimuth angle.
Combining with initial condition $(r, \psi, p_r)|_{t=t_{\rm ini}} = (r_{\rm a}, 0, 0)$ (i.e., starting from the apocenter), 
we obtain the orbital motion $r(t), \psi(t)$.

Assuming the accretion disk lies on the equator $x-y$ plane (the disk angular momentum is in the $z$ direction, 
i.e. $\vec n_{\rm disk} = \vec e_z$), and the orbital plane lies on the $x'-y'$ plane, and the two coordinate frames are related by 
Euler rotations $R_z(\gamma)R_x(\beta)R_z(\alpha)$,
where 
\be
R_z(\alpha) =
\begin{bmatrix}
  \cos\alpha &   -\sin\alpha & \\
  \sin\alpha &   \cos\alpha  & \\ 
   &  & 1
\end{bmatrix} \ ,
\ee
and 
\be
R_x(\beta) =
\begin{bmatrix}
1 & & \\
 &  \cos\beta &   -\sin\beta \\
 &  \sin\beta &   \cos\beta   
\end{bmatrix} \ .
\ee
The orbital motion in the two frames are 
\be 
(x', y', z') = r(\cos\psi, \sin\psi, 0)\ ,
\ee 
and 
\be 
(x, y, z)^{\rm T} = R_z(\gamma)R_x(\beta)R_z(\alpha) (x', y', z')^{\rm T}\ ,
\ee 
respectively. Specifically, $\beta$ is the angle between the disk plane and the orbital plane,
i.e., $\iota_{\rm sd} = {\rm min.}\{\beta, \pi-\beta\}$. 

Using a coordinate frame such that the line of sight (los) lies in the $x-z$ plane with $\vec n_{\rm los} = (\sin\theta_{\rm los}, 0, \cos\theta_{\rm los})$.
The observable collisions happen when $z(t)=\pm H$, where the $\pm$ sign depend on the observation is on the upper/lower half plane. The propagation times of different flares at different collision locations $r_{\rm crs} \vec n_{\rm crs}$
to the observer will also be different.
We can write $t_{\rm obs} = t_{\rm crs} + \delta t_{\rm geom} + \delta t_{\rm shap}$, where 
\be\label{eq:tobs}
\begin{aligned}
    \delta t_{\rm geom} &= -r_{\rm crs} \vec n_{\rm los}\cdot \vec n_{\rm crs}\ , \\ 
    \delta t_{\rm shap} &=-2M_\bullet\ln[r_{\rm crs}( 1+ \vec n_{\rm los}\cdot \vec n_{\rm crs})]\ ,
\end{aligned}
\ee 
are corrections caused by different path lengths and different Shapiro delays \cite{Shapiro1964}, respectively.

To summarize, there are $8$ parameters in the flare timing model: the intrinsic orbital parameters $(a, e)$,
the extrinsic orbital parameters $(\alpha, \beta, \gamma)$, the 
los angle $\theta_{\rm los}$, the  time $t_{\rm ini}$ at apocenter right before the 1st flare observed, 
and the mass of the SMBH $M_\bullet$ or equivalently the Newtonian orbital period $T_{\rm obt} :=2\pi (a/M_\bullet)^{3/2} M_\bullet$. Without loss of generality, we set $\theta_{\rm los} \leq \pi/2$, i.e., the observer locates in the upper half plane, and the observed flares start when the EMRI crosses the upper surface of the disk $z=H$ (we set the disk thickness as $H=1.5 M_\bullet$).

\section{Applying the EMRI+TDE disk model to GSN 069  QPEs}\label{sec:application}
For a given EMRI system and a TDE accretion disk, we can in principle predict the collision times between 
the SMO and the disk, the initial condition of plasma ball from each collision, and the resulting QPE light curve.
Both the EMRI orbital parameters and the disk model parameters can be constrained simultaneously from the QPE light curves.
In this work, we choose to constrain the EMRI kinematics and the plasma ball emission separately: we first fit each QPE with the flare model and obtain the starting time of each flare $t_0\pm \sigma(t_0)$, 
which is identified as the observed disk crossing time. 
In this way, the EMRI kinematics is minimally plagued by the uncertainties in the disk model because the disk crossing time inferred from the QPE light curve is not expected to be sensitive to the disk model.

According to the Bayes theorem, the posterior of parameters given data is
\be \mathcal P(\mathbf{\Theta}|d) \propto \mathcal{L}(d|\mathbf{\Theta}) \pi(\mathbf{\Theta})\ , \ee
where $\mathcal{L}(d|\mathbf{\Theta})$ is the likelihood of detecting data $d$ given a model with parameters $\mathbf{\Theta}$ and $\pi(\mathbf{\Theta})$ is the parameter prior assumed. For the emission model, the likelihood is defined as 
\be 
 \mathcal{L}_{\rm emission}(d|\mathbf{\Theta}) = \prod_{i}\frac{1}{\sqrt{2\pi (F\sigma_i)^2}}
 \exp\left\{-\frac{(L(t_i)-d_i)^2}{2(F\sigma_i)^2} \right\} \ ,
\ee 
where $d_i, \sigma_i$ are the measured QPE luminosity and errorbar at $t_i$, respectively,  
$L(t_i) = L_X(t_i) + L_{\rm bgd}(t_i)$ is the model predicted luminosity [Eqs.~(\ref{eq:model},\ref{eq:bgd}], and $F$ is a scale factor taking possible calibration uncertainty into account. Therefore  we have model parameters $\mathbf{\Theta}=\{
t_0, R_0, T_{\rm e0}, \tau_m, v_\mathrm{exp}, B, A, P_{\rm QPO}, \phi_{\rm QPO}, F\}$ for each flare in the expanding plasma ball model, and  $\mathbf{\Theta}=\{
t_0, t_p, \tau_2, L_p, B, A, P_{\rm QPO}, \phi_{\rm QPO}, F\}$ for each flare in the phenomenological model.
The posterior $t_0\pm \sigma(t_0)$ of each flare is then fed into the flare timing model.

For the flare timing model, the likelihood is  defined in a similar way as 
\be 
 \mathcal{L}_{\rm timing}(d|\tilde{\mathbf{\Theta}}) = \prod_{k}\frac{1}{\sqrt{2\pi (F_t\sigma(t_0^{(k)}))^2}}
 \exp\left\{\frac{(t_{\rm obs}^{(k)}- t_0^{(k)})^2}{2(F_t \sigma(t_0^{(k)}))^2} \right\} \ ,
\ee 
where $t_{\rm obs}^{(k)}$ is the model predicted starting time of $k$-th flare in the observer's frame [Eq.~(\ref{eq:tobs})], 
$t_0^{(k)}$ and $\sigma(t_0^{(k)})$ are the flare starting time and the uncertainty of $k$-th flare (see Table~\ref{tab:t_0_P_QPO}).
In the same way we have 
included a scale factor $F_t$ taking possible systematics/unmodeled physics processes into consideration.
Therefore we have parameters 
$\tilde{\mathbf{\Theta}} = \{a, e, \alpha, \beta, \gamma, T_{\rm obt}, t_{\rm ini}, \theta_{\rm obs}, F_t\}$ 
in the flare timing model. The model parameter inferences are performed using the \texttt{dynesty} \cite{Speagle2020} and \texttt{nessai} \cite{nessai} algorithm 
in the package \texttt{Bilby} \cite{Ashton2019}.

\subsection{GSN 069 QPE light curves}

\begin{figure*}
\includegraphics[scale=0.52]{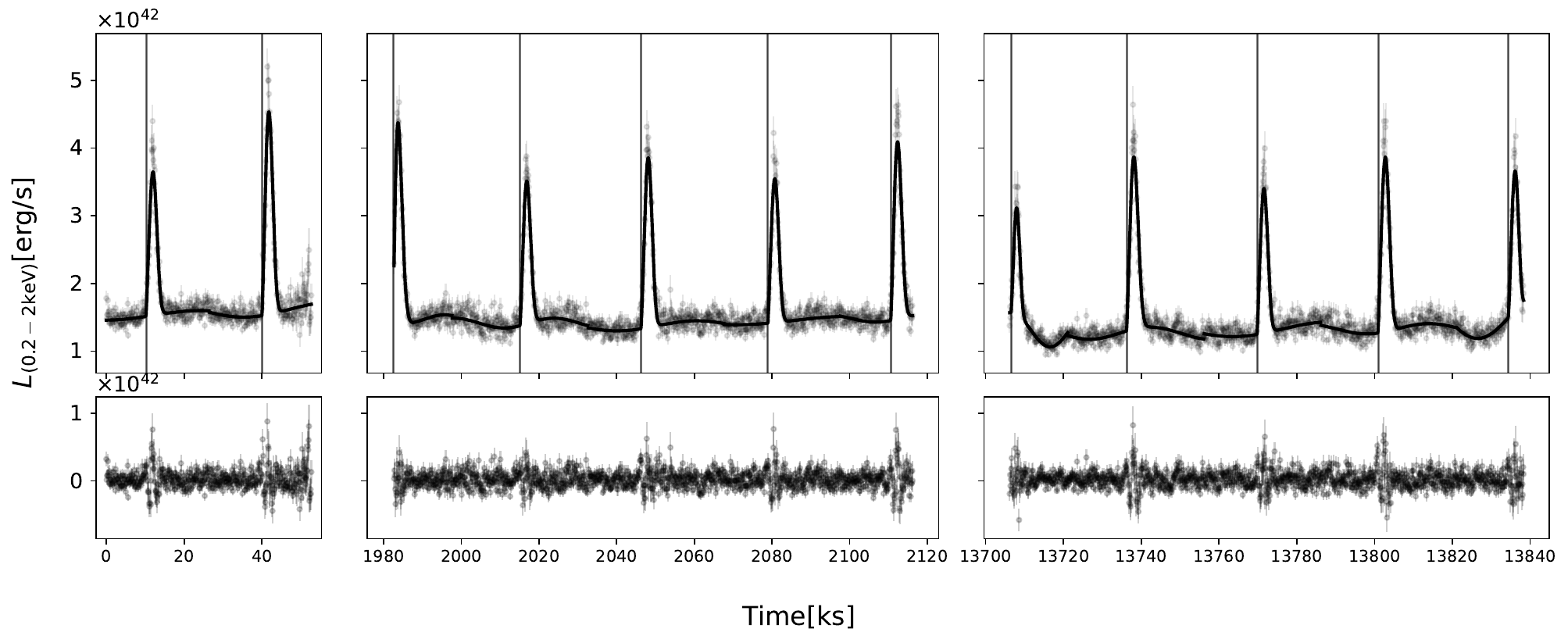}
\caption{\label{fig:lc_fit} Top panel: All the light curve data from XMM3-5 and the best fit results (solid lines).  Vertical bands denote the flare starting times $t_0$. Bottom panel: the residues of the best fit.  We have changed the reference time to the starting time of XMM3 observation.}
\end{figure*}

\begin{figure*}
\includegraphics[scale=0.52]{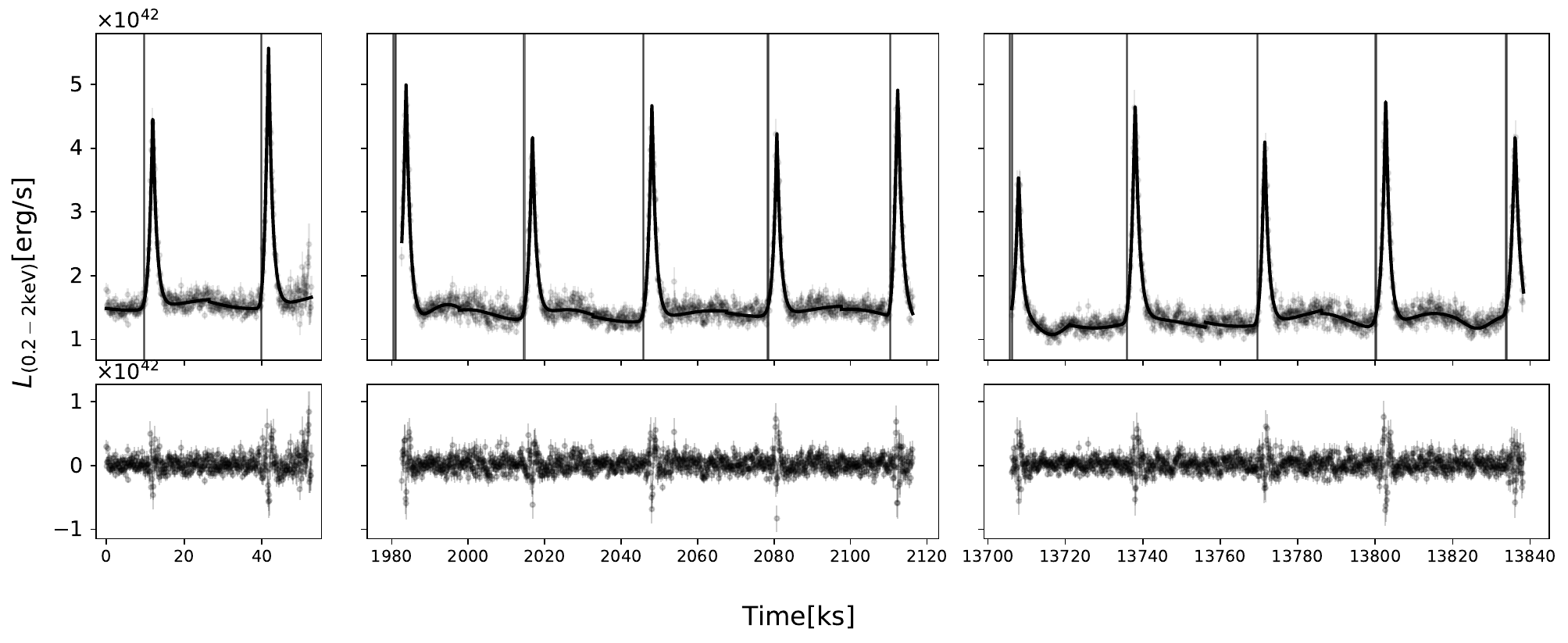}
\caption{\label{fig:lc_fit_phenom} Same as Fig.~\ref{fig:lc_fit}, but for the phenomenological model.}
\end{figure*}

GSN 069 is the first QPE source discovered, and has been monitored extensively in the past decade, including XMM 1-12 and Chandra \cite{Miniutti2023,Miniutti2023b}. From the quiescent state light curves, it is likely that two (partial) TDEs has happened. QPEs are found only in XMM 3-6 and 12 and the Chandra observation, when the quiescent state luminosity is low. 

We reprocessed the raw data from EPIC-pn camera \cite{EPIC-pn} of the XMM-Newton mission, using the latest XMM–Newton Science Analysis System (SAS) and the Current Calibration Files (CCF). The photon arrival times are all barycentre-corrected in the DE405-ICRS reference system. The count-rate-luminosity conversion is accomplished with XSPEC \cite{Xspec}. The result light curves are shown in Fig.~\ref{fig:raw_lc}.\footnote{We did not include the Chandra observation at 14 Feb 2019 in our analysis, 
because the data quality is much lower compared with that in XMM-Newton observations, 
especially the discontinuous features  in the count rates around the background level
make it hard to pin down the flare starting times. This is attributed to the much smaller effective area of Chandra.}

During XMM 3-5 and Chandra (regular phase), the quiescent state luminosity is slowly declining and  the QPEs show clear long-short pattern in the occurrence times and strong-weak pattern in the intensities. During XMM 6, 
the  quiescent state luminosity is climbing up probably in the rising phase of the second TDE 
and the QPEs become irregular in the sense that the alternating strong-weak pattern is not well preserved,
where the first flare does not fit in either the strong or the weak ones, though flares 2-4 still follow 
the alternating long-short and strong-weak patterns (see Fig.~\ref{fig:raw_lc}).
During the recent observation epoch XMM 12, only two QPEs were detected limited by the short exposure time and it is unclear
whether the QPEs have settled down to a new regular phase or not.

From Fig.~\ref{fig:raw_lc}, it is evident that there is an alternating long-short pattern in the QPE recurrence times 
with $T_{\rm long}-T_{\rm short}\approx 2\sim 8$ ks, and $T_{\rm long}+T_{\rm short}$ is approximately a constant.
To quantify these features, we fit these QPEs with the two flare models.
In Fig.~\ref{fig:lc_fit}, we show the 0.2-2 keV light curves of  GSN 069 QPEs from XMM 3-5 along with the best-fit emission model (see the corner plot of the model parameter posterior in Fig.~\ref{fig:eg_corner}). From the residue plots in the lower panels, we see that the fits are reasonable 
for the majority parts of the light curves except around the peaks where the 
sharp turnovers are not captured by the fits, and around the flare staring times $t_0$ where 
the precursor-like features prior to the main flares cannot be captured either.
These limitations also motivate us to consider the alternative phenomenological light curve model (Eq.~\ref{eq:phen}). In Fig.~\ref{fig:lc_fit_phenom}, we show the results of the best-fit phenomenological model,
which largely improves the residues around the light curve peaks and yield consistent 
flare starting times $t_0$ with those from the plasma ball model. The standing-out residues around the 
flare starting times imply possible processes that are not modeled in either model 
(see more discussion in Appendix~\ref{app:hd_sim} where we conduct hydrodynamic simulations of SMO-disk collisions 
trying to identifying the unmodeled processes by
comparing the simulations with the light curves).

\begin{table}
    \centering
    \resizebox{1.0\columnwidth}{!}{%
    \begin{tabular}{r|ll|lll}
       Plasma ball model & $t_0$ & $P_{\rm QPO}$ & $T_{\rm short}$& $T_{\rm long}$& $T_{\rm sum}$ \\
        \hline
          XMM3  flare 1 & $662030843^{+91}_{-86}$ & $51^{+8.53}_{-9.91}$ & $29820^{+123}_{-126}$ & & \\ 
         2 & $662060666^{+67}_{-62}$ & $48^{+9.24}_{-11.39}$ & & &\\
      XMM4  flare 1 & $664003075^{+105}_{-82}$ &  $34^{+5.84}_{-7.17}$ & $31448^{+99}_{-98}$ & $32648^{+96}_{-99}$ & $64092^{+74}_{-68}$\\
        2 & $664035779^{+87}_{-89}$ & $26^{+1.18}_{-1.07}$\\
        3 & $664066973^{+89}_{-85}$ & $44^{+6.26}_{-4.20}$\\
        4 & $664099564^{+84}_{-75}$ & $53^{+4.87}_{-6.98}$\\
        5 & $664131264^{+92}_{-87}$ & $32^{+6.54}_{-6.95}$\\
        
      XMM5  flare 1 & $675727215^{+188}_{-130}$ &$21^{+3.44}_{-2.31}$ & $30393^{+124}_{-132}$ & $33519^{+104}_{-101}$ & $63915^{+92}_{-106}$\\
         2 &   $675756904^{+90}_{-87}$ &$31^{+1.65}_{-1.41}$\\
         3 & $675790441^{+79}_{-78}$ &$52^{+7.48}_{-7.67}$\\
         4 & $675821554^{+105}_{-98}$ &$33^{+2.29}_{-1.84}$ \\
         5 & $675855051^{+83}_{-78}$ &$43^{+11.41}_{-10.75}$ \\
         \hline 
    XMM6 flare 1 & $695049958^{+154}_{-142}$ & $43^{+1.70}_{-1.37}$ & $28567^{+260}_{-245}$ & $36382^{+236}_{-255}$ & $64935^{+250}_{-249}$\\ 
    2 & $695076440^{+132}_{-161}$ & $13^{+0.76}_{-0.62}$\\ 
    3 & $695104985^{+163}_{-141}$ & $30^{+4.48}_{-2.76}$\\ 
    4 & $695141378^{+132}_{-166}$ & $32^{+6.70}_{-7.23}$\\ 
   
    XMM12 flare 1 & $773633586^{+181}_{-210}$ & $61^{+7.31}_{-7.56}$ & $20069^{+223}_{-196}$\\ 
    2 & $773653655^{+74}_{-71}$ & $27^{+3.78}_{-2.32}$
    \end{tabular} }
    \caption{The median values and 1-$\sigma$ uncertainties of the flare starting times $t_0$ [s],
    the QPO periods $P_\mathrm{QPO}$ [ks] and the intervals $T_{\rm short, long, sum}$ [s]
    of GSN 069 QPEs in XMM 3-6 assuming the plasma ball model. }
    \label{tab:t_0_P_QPO}
\end{table}

\begin{table}
    \centering
    \resizebox{1.0\columnwidth}{!}{%
    \begin{tabular}{r|ll|lll}
       Phenomen model & $t_0$ & $P_{\rm QPO}$ & $T_{\rm short}$& $T_{\rm long}$& $T_{\rm sum}$ \\
        \hline
          XMM3  flare 1 & $662030429^{+104}_{-114}$ & $51^{+7.50}_{-8.91}$ & $30025^{+150}_{-158}$ & & \\ 
        2 & $662060452^{+78}_{-82}$ & $52^{+6.80}_{-10.37}$ & & &\\
      XMM4  flare 1 & $664001278^{+397}_{-326}$ & $34^{+8.88}_{-8.08}$ & $31589^{+154}_{-146}$ & $33243^{+236}_{-252}$ & $64842^{+186}_{-210}$\\
        2 & $664035220^{+158}_{-180}$ & $26^{+1.19}_{-1.03}$\\
        3 & $664066491^{+96}_{-97}$ & $45^{+6.50}_{-4.15}$\\
        4 & $664099063^{+136}_{-163}$ & $50^{+8.15}_{-8.65}$\\
        5 & $664130961^{+73}_{-74}$ & $51^{+7.19}_{-10.46}$\\
        
      XMM5  flare 1 & $675726607^{+357}_{-331}$ & $18^{+9.23}_{-2.51}$ & $30258^{+217}_{-215}$ & $33651^{+152}_{-154}$ & $63895^{+221}_{-201}$\\
        2 & $675756547^{+86}_{-95}$ & $33^{+2.54}_{-1.96}$\\
        3 & $675790212^{+81}_{-86}$ & $55^{+4.96}_{-6.66}$\\
        4 & $675820783^{+148}_{-162}$ & $31^{+1.74}_{-1.43}$\\
        5 & $675854426^{+171}_{-191}$ & $16^{+5.66}_{-2.70}$\\
        \hline 
        XMM6 flare 1 & $695049346^{+220}_{-224}$ & $43^{+1.38}_{-1.22}$ & $28345^{+983}_{-919}$ & $36847^{+491}_{-505}$ & $65175^{+947}_{-881}$\\ 
        2 & $695075342^{+790}_{-783}$ & $18^{+2.27}_{-5.41}$\\ 
        3 & $695103697^{+352}_{-341}$ & $26^{+1.61}_{-1.34}$\\ 
        4 & $695140539^{+264}_{-262}$ & $38^{+24.37}_{-25.53}$\\ 
 
    XMM12 flare 1 & $773633107^{+245}_{-268}$ & $54^{+5.23}_{-5.05}$& $20267^{+300}_{-260}$\\ 
    2 & $773653374^{+83}_{-87}$ & $39^{+13.81}_{-8.84}$
    \end{tabular} }
    \caption{Same as Table~\ref{tab:t_0_P_QPO}, but for the phenomenological model.}
    \label{tab:t_0_P_QPO_phenom}
\end{table}

In Table~\ref{tab:t_0_P_QPO} and Table~\ref{tab:t_0_P_QPO_phenom}, we list all the flare starting times $t_0$, the QPO periods $P_{\rm QPO}$ and the intervals $T_{\rm short, long, sum}$ fitted from the QPE light curves with the plasma ball model and the phenomenological model respectively.
With the flare starting times $t_0^{(k)}$, we are to quantify the long-short pattern in the QPE recurrence times.
In XMM 3, 2 flares are observed, therefore only the short time $T_{\rm short}$ can be calculated.
In XMM 4, 5 flares are observed, and  we define
\be 
\begin{aligned}
    T_{\rm long} &= \frac{1}{2}\left(t_0^{(2)}-t_0^{(1)} + t_0^{(4)}-t_0^{(3)} \right)_{\rm  XMM 4} \ , \\
    T_{\rm short} &= \frac{1}{2}\left(t_0^{(3)}-t_0^{(2)} + t_0^{(5)}-t_0^{(4)} \right)_{\rm  XMM 4}  \ , \\
    T_{\rm sum} &= T_{\rm long} + T_{\rm short}\ ,
\end{aligned}
\ee 
and the definitions are similar in XMM 5. 
In XMM 6, 4 flares are observed with the 1st being an outlier, and we define 
\be 
\begin{aligned}
   T_{\rm long} &= \left(t_0^{(4)}-t_0^{(3)} \right)_{\rm  XMM 6} \ , \\
   T_{\rm short} &= \left(t_0^{(3)}-t_0^{(2)} \right)_{\rm XMM 6}  \ , \\
   T_{\rm sum} &= T_{\rm long} + T_{\rm short}\ .
\end{aligned}
\ee 

The measured time intervals $P_{\rm QPO}$ and $T_{\rm short, long, sum}$ are consistent with each assuming 
two different light curve models.
To quantify the evolution of period $T_{\rm sum}$, we fit the $T_{\rm sum}(t)$ with a linear relation and we find 
the slope/the change rate $\dot T_{\rm sum}$ is consistent with zero as 
\be\label{eq:Tsumdot}
\dot T_{\rm sum} = \begin{cases}
(1.5\pm 2.6)\times 10^{-5}  &\text{plasma ball model} \\ 
(-4.7\pm 6.7)\times 10^{-5}  &\text{phenomen model}
\end{cases}
\ee
at $1$-$\sigma$ confidence level.

\begin{figure*}
\includegraphics[scale=0.58]{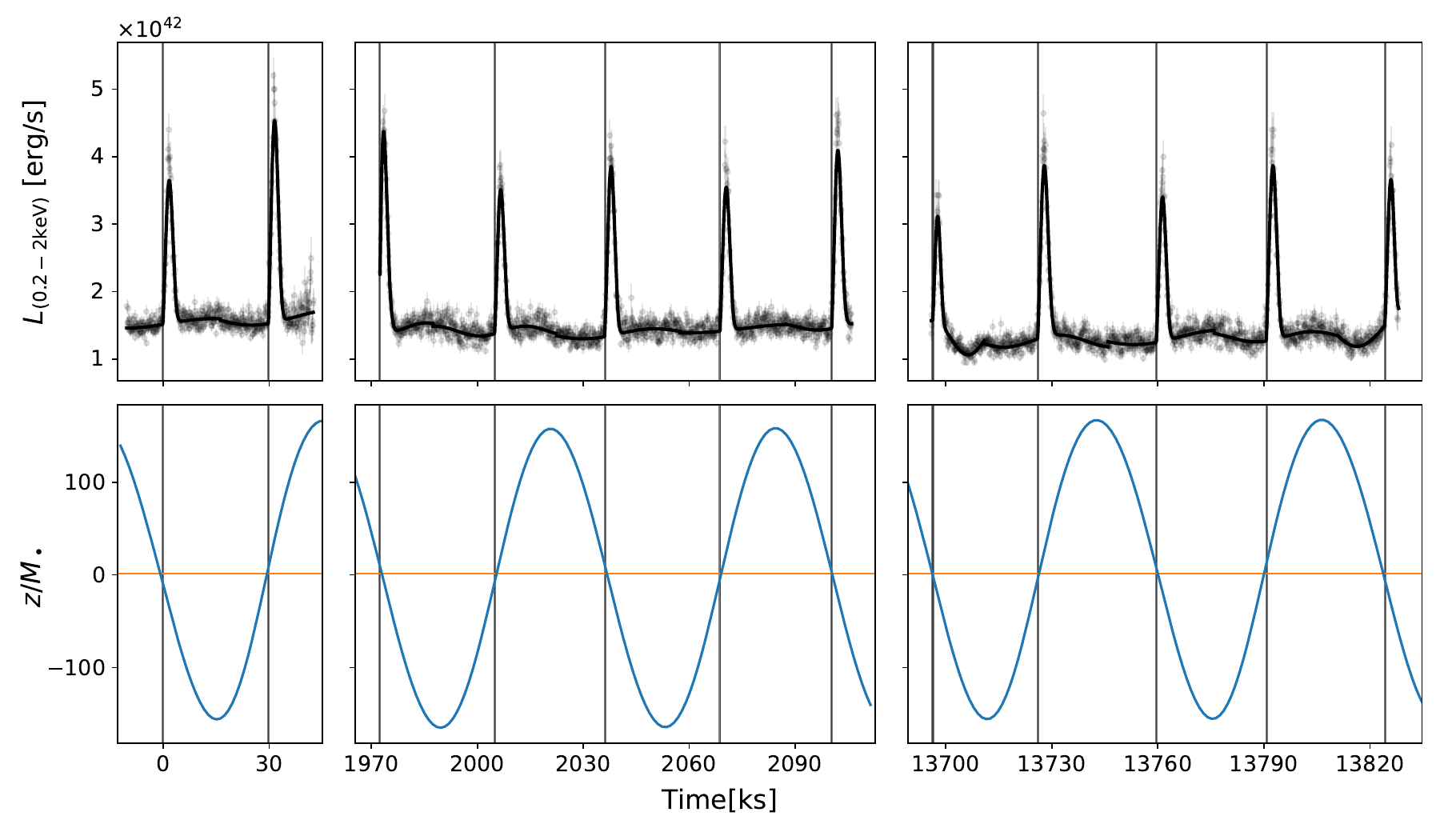}
\caption{\label{fig:lc_orbit} Top panel: light curve data along with the best fit of the emission model, where the vertical lines are the starting times of the QPEs. Bottom panel: $z(t)$ of the best-fit orbit ($a=171 M_\bullet, e=0.04,
T_{\rm obt}= 63.7\ {\rm ks}$), where the horizontal line denotes the disk surface $z=H$.}
\end{figure*}

\subsection{EMRI orbital parameters}

As explained in the previous subsection, QPEs found in XMM 3-5  are in the regular phase and we will 
focus on the QPE timing in these 3 observations for our EMRI orbit analysis (see Summary section
for more discussions about observations XMM 6 and 12).

In Fig.~\ref{fig:lc_orbit}, we show the EMRI orbit of the best-fit flare timing model
(with orbital parameters  $a=171 M_\bullet, e=0.04,
T_{\rm obt}= 63.7\ {\rm ks}$, and the corresponding SMBH mass $M_\bullet=1.05\times 10^6 M_\odot$), 
along with the starting time $t_0$ of each flare 
(Table~\ref{tab:t_0_P_QPO}). In this model, $T_{\rm sum}$ is simply the orbital period $T_{\rm obt}$ which is a constant if the TDE disk lies on the equator and stays in a steady state as assumed, while 
the alternating recurrence times are the result of a non-circular orbit plus the different 
photon propagation times to the observer,
and the variation of $T_{\rm long, short}$ is the result of the SMO orbital precession.

From Fig.~\ref{fig:lc_orbit}, we also see that the brighter peak is sometimes associated with the ascending node,
and sometimes with the descending node of the orbit. The reason is that the QPE luminosity is mainly determined by the relative velocity between the SMO and the disk at the collision point, and the local gas density. 
From simulations in \cite{Ivanov1998} and the HD simulations shown in  Appendix \ref{app:hd_sim},  the amount of gas bursting out from both sides of the disk after a collision are found to be roughly symmetric, therefore the collision direction plays a minor role in the QPE luminosity. The collision location and the collision velocity will be modulated on the SMO orbital precession timescale, therefore
the QPE intensities will be modulated on the same timescale as explicitly shown in \cite{Franchini2023}.

The posterior corner plot of all the model parameters is shown in Fig.~\ref{fig:obt_corner} in Appendix \ref{app:corner},
where the orbital parameters are constrained as 
\be\label{eq:obt_pars}
\begin{aligned}
    a&= 311^{+100}_{-181} M_\bullet,\\ 
    e&=0.04^{+0.04}_{-0.03}, \\
T_{\rm obt}&= 63709\pm 5\ {\rm sec}\ ,
\end{aligned}
\ee 
at 2-$\sigma$ confidence level, respectively.
The constraint of the SMBH mass is therefore $M_\bullet \in (2.4\times 10^5, 1.4\times 10^6) M_\odot$
at 2-$\sigma$ confidence level. In fact, the flare timing contains more information of the SMBH mass than 
what the simple confidence level shows: there are 3 peaks in the posterior of the semi-major axis 
locating at $a=\{160, 220, 330\} M_\bullet$, respectively, which corresponding to 3 favored 
values of the SMBH mass $M_\bullet=\{10^6, 6.2\times 10^5, 3.4\times 10^5\} M_\odot$.

For comparison, we did the similar analysis using the flare starting times $t_0^{(k)}$ in Table~\ref{tab:t_0_P_QPO_phenom} and the posterior corner plot of model parameters is shown in Fig.~\ref{fig:obt_corner_phen}, which is completely consistent with the result shown in Fig.~\ref{fig:obt_corner}.

\section{Summary and discussion}\label{sec:summary}

In this section, we first summarize different aspects of QPE observations 
in addition to the recurrence timescales ($\sim$ a few hours) and the luminosity magnitudes of QPE flares in soft X-rays ($\sim 10^{42}$ ergs/s), and evaluate the performance of the EMRI+TDE disk model in interpreting these observations.
Based on the orbital parameters inferred from the flare timing, we then examine which formation channel the EMRI may come from.  We conclude this section with a brief discussion about possible ways to distinguish the two and limitations of the flaring 
time model used in this paper.

\subsection{Observations versus model predictions}

A reasonable model must be able to naturally explain or at least be compatible with the observations.
In the EMRI+TDE disk model, there are two components in need: a TDE disk and a SMO in a proper position.
In this subsection, we examine whether the model can naturally explain the QPE observations.

1. \emph{Alternating long-short occurrence times:}  in all the confirmed QPE sources to date with more than 3 flares detected (GSN 069,
eRO-QPE1, eRO-QPE2, and RX J1301.9+2747), QPEs have two alternating occurrence times $T_{\rm long}$ and $T_{\rm short}$ \cite{Miniutti2019,Giustini2020,Arcodia2021,Arcodia2022}. 
In the EMRI+TDE disk model, the secondary SMO crosses the accretion disk twice per orbit and two different occurrence times 
$T_{\rm long}$ and $T_{\rm short}$ alternate as a result of
non-circular orbit and different delays of two consecutive collisions [see Eq.~(\ref{eq:tobs})].  (\checkmark)

2. \emph{Alternating strong-weak QPE intensities:}  strong-weak QPEs alternately occur \cite{Miniutti2019,Giustini2020,Arcodia2021,Arcodia2022,Chakraborty2021}.  
In the EMRI+TDE disk model, the secondary BH crosses the accretion disk twice per orbit producing two 
different flares because  the two disk crossings 
are not identical to each other depending on the orbit eccentricity. (\checkmark)

3. \emph{Spectral evolution:} QPEs  measured in higher energy bands are stronger, peak earlier, and have shorter
duration than when measured at lower energies \cite{Miniutti2019,Giustini2020,Arcodia2021,Arcodia2022,Chakraborty2021,Miniutti2023b}. 
The QPE spectral evolution perfectly matches the flare model prediction where the QPE emission 
comes from an expanding and cooling plasma ball. (\checkmark)

4. \emph{Light curve profile:} a common feature of QPEs is the fast rise and slow decay light curve profile with 
a low QPE duty cycle (a few percent)
\cite{Miniutti2019,Giustini2020,Arcodia2021,Arcodia2022,Chakraborty2021}. Similar to supernova explosions, the thermal radiation from a freely expanding and cooling plasma ball naturally produces the fast rise and slow decay light curve. (\checkmark)

5. \emph{Association with TDEs:} two QPE sources (GSN 069 and XMMSL1 J024916.6-04124) and a candidate (AT 2019vcb), have been directly associated with X-ray TDEs \cite{Shu2018,Sheng2021,Chakraborty2021,Miniutti2023,Quintin2023}. 
In the EMRI+TDE disk model, two components are in need for producing QPEs: a TDE disk and a SMO in a proper position. Therefore we do not expect  to see QPEs in all TDEs. The fraction depends on the availability of the 
second component.
From the QPEs that are associated with TDEs, we may statistically infer the distribution 
of SMOs around SMBHs, and therefore the EMRI formation rate. In this aspect, QPEs might be a unique probe to 
EMRIs in the pre-LISA era. (\checkmark)

6. \emph{Association with past AGN activities, and not with on-going AGNs:}
the presence of  narrow lines in all QPE host galaxies and 
the absence of luminous broad emission lines indicates that they are recently switched-off  AGNs
\cite{Wevers2022,patra2023} (but see also \cite{Metzger2022} for a different interpretation). 
In on-going AGNs, the majority of SMOs in the vicinity of the SMBH are those captured onto the AGN accretion disk \cite{Pan2022}, 
as a result the number of SMOs crossing the AGN disk is suppressed,
therefore no QPE association with on-going AGNs. 
Past AGN activities are expected to boost the EMRI formation by accelerating the inward migration of SMOs
(known as the wet EMRI formation channel \cite{Pan2021,Pan2021prd,Pan2021b,Pan2022}).
In recently turned-off AGNs, SMOs are accumulated on the equator of the SMBH with $a \sim \mathcal{O}(10^2) M_\bullet$  (see \cite{Pan2022,Pan2021} for full Fokker-Planck calculation or 
Section~\ref{sec:implication} for a sketch),
while the accretion disk formed from a TDE is in general not exactly aligned with the equator, therefore QPEs are produced as the SMO crossing the TDE formed accretion disk. 
As we will show in Section~\ref{sec:implication}, the GSN 069 EMRI orbit is consistent with the wet channel prediction,
while incompatible with alternatives. This adds another independent evidence for the past AGN activities in GSN 069.
(\checkmark)

7. \emph{(Anti-)association with SMBH mass:} QPEs are preferentially found in nuclei of dwarf galaxies, where the SMBHs 
are relatively light with mass no more than a few times $10^6 M_\odot$ \cite{Wevers2022,Miniutti2023}. 
In the EMRI+TDE disk model, this anti-association with the SMBH mass comes from the TDE rate dependence on the SMBH mass,
and the finite size of the TDE formed accretion disk.
TDEs have been found to be preferentially around lighter SMBHs (see e.g., \cite{Wang2004,Stone2016}).
Assuming an $\alpha$-disk of total mass $m_{\rm d}$, we have the disk size
\be 
r_{\rm d}  = 10^2 M_\bullet \left(\frac{m_{\rm d}}{0.35 M_\odot} \frac{\alpha}{0.01}\frac{\dot M_\bullet}{0.1 \dot M_{\bullet,\rm Edd}}  \right)^{2/7} \left(\frac{M_\bullet}{3\times10^6 M_\odot}\right)^{-4/7}\ .
\ee 
For more massive SMBHs, the disk  is smaller, and $r_{\rm d}/M_\bullet < 10^2$ for a SMBH heavier than $\sim 3\times10^6 M_\odot$.
The chance of SMOs with  $a \sim \mathcal{O}(10^2) M_\bullet$ crossing the TDE formed disk around a more massive SMBH is lower.
(\checkmark)

8. \emph{(Anti-)association with SMBH accretion rate in quiescent state:} 
long-term observation of GSN 069 shows that QPEs may only be present below a quiescent luminosity threshold $L_{\rm disk, bol}\sim 0.4 L_{\rm Edd}$  \cite{Miniutti2023}.  There are two possible origins of this anti-association: 
the ratio of the QPE luminosity to the disk luminosity in the same energy band $L_{\rm QPE}/L_{\rm disk, X}$ depends on the disk accretion rate \cite{Franchini2023}, or QPEs are delayed relative to the TDE by a time interval (which is about a few years in the case of GSN 069).

In the $\alpha$ disk model, the disk surface density $\Sigma \propto \dot M_\bullet^{-1}$ and the energy deposited in the SMO-disk collision $\delta E_{\rm SMO}\propto \Sigma$ [Eqs.~(\ref{eq:delta_E_star},\ref{eq:delta_E})] is therefore lower for the higher accretion rate case. The thermal luminosity from the accretion disk  $L_{\rm disk, bol}\propto \dot M_\bullet$
is higher in the higher accretion case and the dependence is more sensitive for the soft X-ray luminosity $L_{\rm disk, X}$.
As a result,  the QPE is harder to identify from the luminous background in the higher accretion rate case 
(see  \cite{Franchini2023} for the explanation in terms of QPE temperatures and the disk temperature).

The SMO orbit is in general not aligned with the TDE disk. The misaligned disk initially precess like a rigid body before settling down to 
a non-precessing warped disk \cite{Liska2018,Chatterjee2020,Chatterjee2023}. In a recently turned-off AGN, the sBHs are preferentially found on the equator of the SMBH.
As a result, the sBH is expected to be highly inclined with respect to the new TDE disk when the sBH-disk collisions are less energetic [Eq.~(\ref{eq:delta_E})], and QPEs emerge only when the two become nearly aligned and the collisions are sufficiently energetic. This scenario works for sBH EMRIs only, 
where the emergence of QPEs is delayed by the disk alignment timescale, the accurate value 
of which is not accurately calculated \cite{Stone2012,Franchini2016}.
(\checkmark)

9. \emph{Stability:} during the observations XMM 3-5 (regular phase dubbed in  \cite{Miniutti2023}), the QPE occurrence times
$T_{\rm sum} :=T_{\rm long} + T_{\rm short}$
are stable with variation rate $\dot T_{\rm sum}$ consistent with zero (see Eq.~(\ref{eq:Tsumdot})). 
The sBH orbital period change rate due to disk crossing turns out to be  (Eq.~[\ref{eq:delta_E}])
\be\label{eq:Tdot}
\begin{aligned}
    \dot T_{\rm obt} &= \frac{dT_{\rm obt}}{dE_{\rm sBH}} \dot{E}_{\rm sBH}
=3\frac{\delta E_{\rm sBH} }{E_{\rm sBH}} \approx  -3\times 10^{-6}\left(\frac{\ln\Lambda}{10}\right)   \\ 
&\times  \alpha_{0.01}^{-1} \dot M_{\bullet,0.1}^{-1} M_{\bullet,6}^{-7/3}  T_{\rm obt, 20}^{7/3}
\left( \frac{m}{30 M_\odot}\right)  \left(\frac{\sin\iota_{\rm sd}}{0.1}\right)^{-3} ,
\end{aligned}
\ee 
where $T_{\rm obt, 20}:= T_{\rm obt}/20\ {\rm hr}$.
The orbital change rate is consistent with the observed $\dot{T}_{\rm sum}$.
The star orbital period change rate is given by \cite{Linial2023c} 
\be \label{eq:Tdot_star}
\begin{aligned}
    \dot T_{\rm obt} 
    &\approx -2\times 10^{-5} \alpha_{0.01}^{-1} \dot M_{\bullet,0.1}^{-1} M_{\bullet,6}^{-1}  T_{\rm obt, 20}
m_{\star,\odot}^{-1} R_{\star,\odot}^2 \sin\iota_{\rm sd}\ ,
\end{aligned}
\ee 
where $m_{\star,\odot} :=m_\star/m_\odot, R_{\star,\odot} :=R_\star/R_\odot$.
The orbital period change rate is also consistent with the current observation constraint.
(\checkmark)

10. \emph{Association with QPOs in quiescent state: }  in the old phase (XMM 3-5)  
quiescent level QPOs were detected with a period close to the corresponding QPE recurrence times $P_{\rm QPO}\in 
(T_{\rm sum}/4, T_{\rm sum})$ (see Tables~\ref{tab:t_0_P_QPO}, \ref{tab:t_0_P_QPO_phenom}), and a $8 \sim 10$ ks delay with respect to the QPEs  \cite{Miniutti2023,Miniutti2023b}. 
It is unclear whether the periodic EMRI impacts on the accretion disk 
are able to produce the quiescent level QPOs with correct time delay, period and amplitude. \footnote{Previous simulations show that the accretion rate $\dot M_\bullet$ of the central BH is indeed modulated 
by periodic stellar-disk collisions \cite{Sukova2021}.}  (?)

11. \emph{New QPE phase in  GSN 069:} during the old phase (XMM 3-5), both the QPE intensities and occurrence times 
alternated with $T_{\rm long} \approx 33 $ ks , $T_{\rm short} \approx 31 $ ks
and $T_{\rm sum}  \approx 64 $ ks. During the irregular phase (XMM 6) which is on the rise of TDE 2, the QPE occurrence times observed do not follow the alternating patterns exactly with the 1st flare fitting in neither the strong ones nor the weak ones, though flares 2-4 still follow the alternating long-short and strong-weak patterns with $T_{\rm long}^{\rm (irg)} \approx 36$ ks , $T_{\rm short}^{\rm (irg)} \approx 28$ ks and $T_{\rm sum}^{\rm (irg)}  \approx 64$ ks. After disappeared for two years (XMM 7-11), QPEs reappeared (XMM 12) with quite different occurrence times from 
those in the old phase or the irregular phase: $T_{\rm short}^{\rm (new)}\approx 20$ ks  and $T_{\rm long}^{\rm (new)} (> 27\ {\rm ks})$ is not fully resolved because only two QPEs have been detected \cite{Miniutti2023}. By fitting the variation in the quiescent level with a sine function, the period  was constrained to be $P_{\rm QPO}^{\rm (new)} =54\pm 4$ ks, however this period suspiciously coincides with the exposure time and the statistical quality of the fit is poor with reduced $\chi^2_\nu\approx 2.9$ ($\chi^2=140$ and $N_{\rm dof}=49$) \cite{Miniutti2023,Miniutti2023b}. This QPO period has been speculated to be same to the QPE period in the new phase, $P_{\rm QPO}^{\rm (new)} \approx T_{\rm sum}^{(\rm new)}$, 
though this speculated relation is not observed in the previous phases where the QPO periods $P_{\rm QPO}$ are found  in the range of  $(T_{\rm sum}/4, T_{\rm sum})$ (see Tables~\ref{tab:t_0_P_QPO} and \ref{tab:t_0_P_QPO_phenom}). Further measurements with longer exposure time are needed to confirm whether a different QPE period 
 emerges in the new phase. 

In the EMRI+ TDE disk model, the shorter $T_{\rm short}^{\rm (new)}$ is a result of a puffier disk due to the higher accretion rate sourced by the TDE 2
and consequently a larger fraction 
\be 
f_{\rm ebd}\approx\frac{4H/\sin\iota_{\rm sd}}{2\pi r} \approx \frac{1}{3}\times\frac{H}{5M_\bullet}\frac{10^2M_\bullet}{r}\frac{0.1}{\sin\iota_{\rm sd}}\ 
\ee
of the EMRI orbit is embedded in the disk. 
As a result, $\approx 1/3$ orbit stays above the disk which is visible to the observer making a shorter $T_{\rm short}^{\rm (new)}\approx T_{\rm obt}/3$, and the remaining $\approx 2/3$ hides in or below the disk which is invisible.
This geometrical effect of the disk thickness may also contributes to the shorter $T_{\rm short}^{\rm (irg)}$ in XMM 6 which is on the rise
of TDE 2.
(\checkmark)

~\\

To identify the true origin(s) of QPEs, all the existing models should be tested against these observations.
Taking GSN 069 as an example, the alternating long-short recurrence pattern and a stable $T_{\rm long}+T_{\rm short}$ (see Fig.~\ref{fig:raw_lc}) together pose a huge challenge for the single-period models. In these models, 
there is no natural explanation to them without a twofold fine tuning: 
alternating delay-advance in the recurrence times for producing the alternating long-short pattern, and 
cancellation of consecutive delay/advance for producing a stable $T_{\rm sum}$. 
In the EMRI+TDE disk model, the alternating long-short pattern and the constant $T_{\rm long}+T_{\rm short}$ are natural consequences of a non-circular orbit,
and most of the QPE observations summarized above can be quantitatively recovered 
though it is unclear whether quiescent-state QPOs can be naturally generated.

In addition to the common properties share by most QPE sources, GSN 069 is special in the sense that two TDE flares with $\sim 9$ years
apart have been observed. Considering the low TDE rate, the two TDEs are likely two consecutive partial disruptions of a same star.
This is similar to another confident repeating partial TDE (pTDE) AT 2020vdq \cite{Somalwar:2023sml}, where two consecutive TDEs are found with an interval $\sim 3$ years.
The relatively tight orbits are unusual for stars that enter the loss cone driven by two-body relaxation, 
because the TDE rate should be dominated by stars on much wider orbits (see Ref.~\cite{Stone:2020vdg} for a detail review,
and see Fig.~\ref{fig:no_lc} and the discussion therein for a pictorial understanding).
A natural explanation is that the tidally disrupted star comes from a previously tidally disrupted binary (see Ref.~\cite{Cufari:2022szx,Somalwar:2023sml} and Fig.~\ref{fig:no_lc}).

There are other models where the second TDE flare is due to runaway envelope
stripping of the impactor star instead \cite{Linial2023}. In this model, there is no need for the repeating partial TDEs,
and the resumption of QPEs could be due to a surviving stellar core. To distinguish the two models, longer monitoring of GSN 069 
and other QPE sources of shorter periods is necessary: 1) after the second flare,  a large change in the orbital period is expected in the runaway envelope stripping model; 2) in QPE sources of shorter periods the second flare should be common because the runaway envelope stripping is more prone to happen.

\subsection{Implications of QPEs on EMRI formation}\label{sec:implication}
If QPEs are indeed sourced by impacts of SMOs and accretion disks formed from TDEs, the EMRI orbital parameters will be invaluable for inferring the EMRI formation channels. 
These SMOs may be  captured by the SMBH via the (dry) loss-cone channel \cite{Hopman2005,Preto2010,Bar-Or2016,Babak2017,Amaro2018,Broggi2022}, Hills mechanism \cite{Miller2005,Raveh2021} 
or the (wet) AGN disk channel \cite{Sigl2007,Levin2007,Pan2021prd,Pan2021b,Pan2021,Pan2022,Derdzinski2023,Wang2023,Wang2023b}.
The EMRI orbits inferred from the 6 QPE sources are of similar properties, with low eccentricity $e$ and with pericenter distance $r_{\rm p}/M_\bullet = \mathcal{O} (10^2)$ \cite{Franchini2023,Linial2023}. These orbital parameters are roughly consistent with the (wet) EMRIs formed in AGN disks, where the wet EMRIs are expected to be nearly circular, and concentrate
around $r= \mathcal{O}(10^2) M_\bullet$ at the end of an  AGN phase \cite{Pan2021prd,Pan2021b,Pan2021,Pan2022},
while they are distinct from those EMRIs formed via the loss-cone channel, 
where the EMRIs are expected to be highly eccentric ($e\rightarrow 1$) 
and sharply concentrate around $r_{\rm p}=10 M_\bullet$ for  sBH EMRIs and around the tidal radius 
$r_{\rm p}=r_{\rm tidal}=R_\star(M_\bullet/M_\star)^{1/3}$
for stellar EMRIs. A more quantitative analysis for GSN 069 EMRI is outlined as follows. 

In a single-component stellar cluster, the relaxation timescale at radius $r$ due to 
2-body scatterings is \cite{Spitzer1987}
\be 
t_{\rm rlx}(r) = \frac{0.339}{\ln\Lambda}\frac{\sigma^3(r)}{m_\star^2 n_\star(r)}\ ,
\ee 
where $\sigma(r)$ is the local velocity dispersion ($\approx\sqrt{M_\bullet/r}$ within the influence radius $r_{\rm h}$ of the SMBH),
$n_\star(r)$ is the star number density, and $\ln\Lambda\approx10$ is the  Coulomb logarithm.
Using the empirical $M_\bullet-\sigma_\star$ relation \cite{Tremaine2002,Gultekin2009}, the influence radius $r_{\rm h}:=M_\bullet/\sigma_\star^2$ turns out to be 
\be 
r_{\rm h} \approx M_{\bullet,6}^{0.5}\ {\rm pc}\ .
\ee 
For a highly eccentric orbit, the angular momentum only needs to change slightly to make
an order unity difference in the orbit, and the diffusion timescale in the angular momentum in general is
shorter than the relaxation timescale in the energy as $t_J\approx (1-e^2) t_{\rm rlx}$. For comparison, the energy dissipation timescale of the star  due to GW emission is \cite{Peters1964}
\be 
t_{\rm GW} = -\frac{a}{\dot a} \approx \frac{a^4}{M^2_\bullet m_\star}\left(\frac{r_{\rm p}}{a} \right)^{7/2}\ ,
\ee 
where $r_{\rm p}=a(1-e)$ is the pericenter distance.
Assuming the Bahcall-Wolf (BW) density profile $n_\star(r) \propto r^{-7/4}$ \cite{Bahcall1976}, 
one can find the ratio \cite{Linial2023b}
\be \label{eq:tratio}
\frac{t_J}{t_{\rm GW}} \approx \left(\frac{r_{\rm p}}{M_\bullet}\right)^{-5/2}\left(\frac{a}{r_{\rm h}}\right)^{-5/4}\ .
\ee 

\begin{figure}
\includegraphics[scale=0.7]{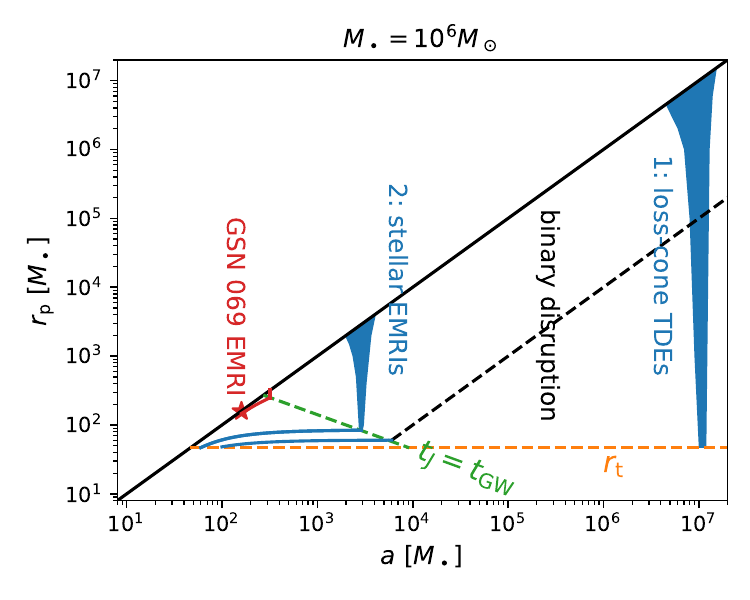}
\caption{\label{fig:no_lc}  Following Ref.~\cite{Linial2023b}, we use the $r_{\rm p}-a$ phase diagram to  analyze the possibility of GSN 069 EMRI formation via
the loss-cone channel and the Hills mechanism. As an example, we consider a $10^6 M_\odot$ SMBH and the tidal disruption radius $r_{\rm t} = (M_\bullet/m_\star)^{1/3} R_\star\approx 47 M_\bullet$ of a solar type star.
In the loss-cone channel, there are in general two fates of the stars: (1) loss-cone TDEs, 
most of them are dominated by 2-body scatterings ($t_J < t_{\rm GW}$) and get tidal disrupted by the SMBH  when scattered into a low-angular momentum orbit
with $r_{\rm p} < r_{\rm t}$, and (2) stellar EMRIs, a small fraction of them are scattered into the GW emission dominated regime ($t_J > t_{\rm GW}$) gradually circularizing and losing mass via partial TDEs.  
Stars residing in a tiny phase space ($r < 300 M_\bullet$ and $e\approx 0$) can possibly become
GSN 069 like EMRIs.
In the Hills mechanism, the bounded star after a binary disruption is highly eccentric with eccentricity $e\approx 0.98$ \cite{Miller2005} (the black dashed line). The star faces the same two fates, a loss-cone TDE or a stellar EMRI,
neither of which end as GSN 069 like EMRIs.}
\end{figure}

To analyze the formation rate of GSN 069 like EMRIs in the loss-cone channel, a $r_{\rm p}-a$ diagram  proposed in Ref.~\cite{Linial2023b} is useful. As shown in Fig.~\ref{fig:no_lc},
most stars are dominated by 2-body scatterings and get tidally disruption when scattered into a 
low-angular momentum orbit with $r_{\rm p} < r_{\rm t}$, a small fraction of stars are scattered into 
the GW emission dominated regime (dubbed as stellar EMRIs) gradually circularising and 
finally lose mass via partial TDEs. The GSN 069 EMRI is one of the stellar EMRIs.
The TDE rate and stellar EMRI formation rate can be estimated as 
\be 
R_{\star}(< r) \approx \frac{N_\star(<r)}{t_{\rm rlx} (r)} \propto r\ ,
\ee 
from which we find the ratio of the formation rate of GSN 069 like EMRIs ( EMRIs with orbital parameters $e<0.1$ and $T_{\rm obt}> 63$ ks that are confidently consistent with that of the GSN 069 EMRI,
and of course in the  GW emission dominated regime with $t_J > t_{\rm GW}$) to the total TDE rate as 
\be 
\frac{R_\star(< 300 M_\bullet)}{R_{\star}(< r_{\rm h}) } \approx 10^{-5}\ ,
\ee 
where we have used the fact that only stars in the range of $r < 300 M_\bullet, e\approx 0$ can
possibly become GSN 069 like EMRIs (see Fig.~\ref{fig:no_lc}).
The ratio becomes even lower if considering the mass-segregation effect where the star density is expected to be suppressed 
at small radii \cite{Hopman2005,Preto2010,Bar-Or2016,Babak2017,Amaro2018,Broggi2022}.
Therefore the stellar EMRI in GSN 069 unlikely comes from the loss-cone channel and similar analysis also applies to sBH EMRIs.

Hills mechanism has been proposed as an efficient EMRI formation channel, and the sBH EMRIs at coalescence from this channel was speculated to be nearly circular due to the long inspiral phase \cite{Miller2005}. However recent simulations taking the mass segregation effect into account show that the orbital eccentricity of sBH EMRIs at coalescence actually peaks at high eccentricity, following a distribution similar to in the loss-cone channel \cite{Raveh2021}. For EMRIs in the earlier inspiral phase with orbital semi-major axis $a=\mathcal{O}(10^2 M_\bullet)$, the orbital eccentricity should be even higher, 
which is in contrast with low-eccentricity EMRI in GSN 069. 
From Fig.~\ref{fig:no_lc}, the same conclusion can be obtained.
In the Hills mechanism, the bounded stars after binary disruptions are highly eccentric with eccentricity $e\approx 0.98$ \cite{Miller2005}. The stars face the same two fates, and neither of them ends as GSN 069 like EMRIs:
most stars end as loss-cone TDEs and the remaining small fractions of stars evolve into stellar EMRIs but they are too eccentric to become GSN 069 like EMRIs. And similar analysis also applies to sBH EMRIs.

\begin{figure}
\includegraphics[scale=0.7]{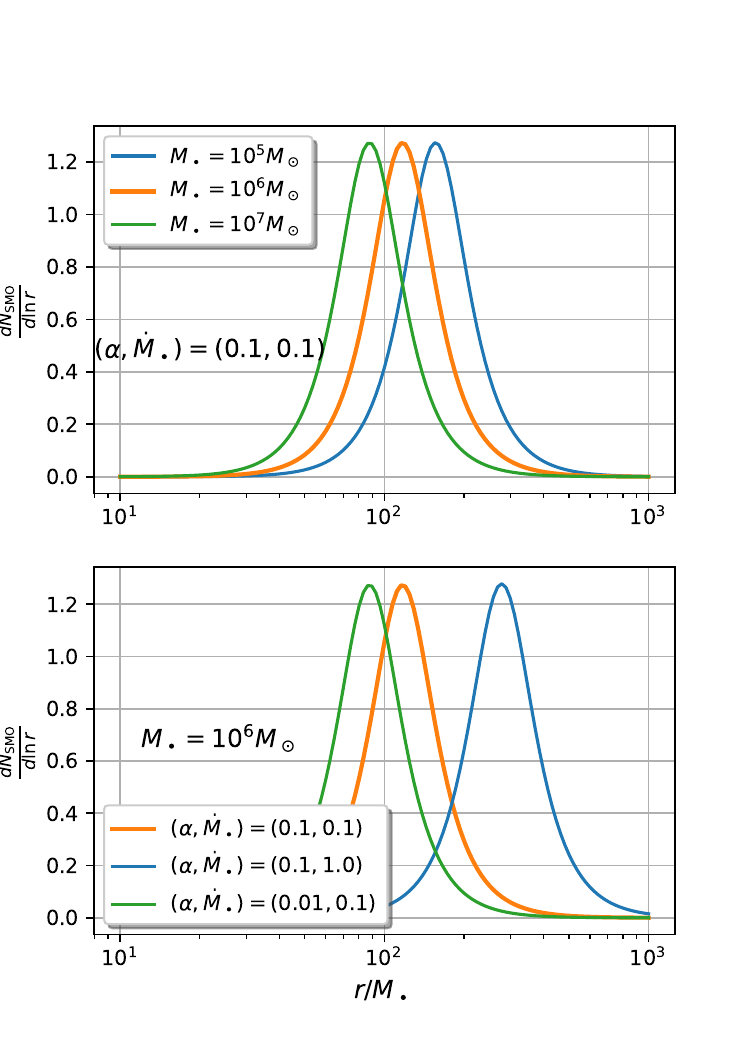}
\caption{\label{fig:sBH_AGN} The distribution $\frac{dN_{\rm SMO}}{d\ln r}$ of SMOs that are captured onto an AGN disk, where we have normalised  the total number $N_{\rm SMO}$ to be unity. Stars follow the same distribution if they are not destroyed in collisions with the AGN disk and with other stars/sBHs.
The AGN disk is modelled as an $\alpha$-disk with viscosity parameter $\alpha$ and  
accretion rate $\dot M_\bullet$ (in units of $\dot M_{\bullet, \rm Edd}$).}
\end{figure}

In the wet EMRI formation channel, a SMO orbiting around an accreting SMBH can be captured to the accretion disk as interactions (dynamical friction and density waves) with the accretion disk tend to decrease to the orbital inclination angle w.r.t. the disk.
After captured onto the disk, the SMO migrates inward driven by the density waves and gravitational wave emission.
The orbital eccentricity is expected to be damped by the density waves to $e\sim h$, where $h$ is the disk aspect ratio.
The number $N_{\rm SMO}(<r)$ of SMO captured is determined by 
\be 
\frac{\partial}{\partial t} N_{\rm SMO} + \frac{\partial}{\partial \ln r} \left( N_{\rm SMO} \frac{v_{r, \rm mig}}{r} \right) = 
F_{\rm cap}(r)\ ,
\ee 
where $F_{\rm cap}(r)$ is the capture rate of SMOs from the nuclear stellar cluster
and $v_{r, \rm mig}$ is the migration velocity in the radial direction. 
The migration is dominated by the GW emission at small separations and by the type-I migration 
at large separations \cite{Pan2021prd}, $v_{r, \rm mig} = v_{r, \rm GW} + v_{r, \rm I}$, with
\be 
v_{r,\rm GW} = -\frac{64}{5}\frac{mM_\bullet^2}{r^3}\ ,
\ee 
and 
\be 
v_{r,\rm I} = -(2.7+1.1\alpha_s)\frac{m\Sigma r^3\Omega}{M_\bullet^2 h^2}\ ,
\ee 
where $\alpha_s := d\ln\Sigma/d\ln r$, 
$\Omega(r)$ is the disk angular velocity,
and $h(r)$ is the disk aspect ratio \cite{Tanaka2002,Tanaka2004}. 
To obtain the SMO number $N_{\rm SMO}(<r)$, SMOs captured by the disk (disk component) and the 
residing the cluster (cluster component) should be evolved self-consistently \cite[see e.g.][]{Pan2021prd,Pan2021b,Pan2021,Pan2022}.
Here we focus on the distribution of disk-component SMOs in the vicinity of a SMBH ($r < 10^3 M_\bullet$), where the 
capture rate is negligible and the distribution is clearly 
$N_{\rm SMO} \propto r/v_{r,\rm mig}$,
in a steady state. As a result, we obtain $N_{\rm SMO} \propto r^4$ in the GW emission dominated regime and 
$N_{\rm SMO} \propto r^{-4}$ in the type-I migration dominated regime for an $\alpha$-disk.
In Fig.~\ref{fig:sBH_AGN}, we show the distributions $dN_{\rm SMO}/d\ln r$ for a few AGN examples, 
where $dN_{\rm SMO}/d\ln r$ in general peaks at $r=\mathcal{O}(10^2 M_\bullet)$ for SMBHs with masses in the range of $10^{5-7} M_\odot$. After the AGN turns off, the SMOs will migrate inward driven solely by GW emissions 
and the distribution of SMOs in the radial direction will the reshaped with the peak moving outwards to a larger radius.
Therefore, the EMRI in GSN 069  is consistent with the 
wet channel expectation.

In the above analysis, we did not take the tidal circularization process of the stellar orbit into account, which turns out to be subdominant. As shown in Ref.~\cite{Verbunt1995}, the  tidal circularization rate of the stellar orbit around a SMBH is 
\be 
\begin{aligned}
    \frac{d\ln e}{dt}\Big|_{\rm tide} 
    &\approx -0.09 T_{\rm eff, 5800}^{4/3}
M_{\rm env, 0.02}^{2/3} M_{\star,\odot}^{-1}
 \left(\frac{M_\bullet}{M_\star} \right)^2 
\left(\frac{R_\star}{a}\right)^8 \ {\rm yr}^{-1}\ , \\ 
&\approx 5\times 10^{-10}  T_{\rm eff, 5800}^{4/3}
M_{\rm env, 0.02}^{2/3} \\ 
&\times \left(\frac{T_{\rm obt}}{64 \ {\rm ks}} \right)^{-16/3} M_{\bullet,6}^{-2/3} M_{\star,\odot}^{-3} R_{\star,\odot}^8
\ {\rm yr}^{-1}\ ,
\end{aligned}
\ee 
where $M_{\rm env}$ is the mass of the stellar convective envelope and $T_{\rm eff}$ is the stellar effective temperature.
In comparison with the circularization rate driven by GW emission \cite{Peters1964}, 
\be 
\begin{aligned}
    \frac{d\ln e}{dt}\Big|_{\rm GW} 
    & = -\frac{304}{15}\frac{M_\bullet^2 M_\star}{a^4} g(e)\ , \\ 
    & \approx -2\times 10^{-7} g(e) \left(\frac{T_{\rm obt}}{64 \ {\rm ks}} \right)^{-8/3}  M_{\bullet,6}^{2/3} M_{\star, \odot} \ {\rm yr}^{-1}\ ,
\end{aligned}
\ee 
where 
\be 
g(e) = (1-e^2)^{-5/2} \left(1+\frac{121}{304}e^2 \right)\ ,
\ee 
we find 
\be 
\begin{aligned}
    \frac{\dot e \ |_{\rm GW}}{\dot e\ |_{\rm tide}}
    &\approx 400  T_{\rm eff, 5800}^{-4/3}
M_{\rm env, 0.02}^{-2/3}\left(\frac{T_{\rm obt}}{64 \ {\rm ks}} \right)^{8/3} 
\\ 
&\times M_{\bullet,6}^{4/3}  R_{\star,\odot}^{-8} 
M_{\star,\odot}^4 g(e)\ .
\end{aligned}
\ee 
Therefore the tidal circularization never was dominant compared with the GW emission and can be safely ignored.

For the SMO settling down to a low-eccentricity orbit around the central SMBH, 
Kozai-Lidov oscillations driven by a third body must be quenched, e.g., by the apsidal precession of the SMO.
The quench condition that the precession period is shorter than the 
Kozai-Lidov oscillation period $P_{\rm pre} < P_{\rm K-L}$ has been derived as \cite{Blaes2002}
\be 
\frac{a_3^3}{a^3} > \frac{3 m_3 a (1-e_3^2)^{3/2}}{4 M_\bullet^2 (1-e^2)^{3/2}}\ ,
\ee 
where $m_3, a_3$ and $e_3$ are the mass, the semi-major axis and the orbital eccentricity of the third body.
For the EMRI system in GSN 069 with $a=\mathcal{O}(10^2) M_\bullet$, the above condition is guaranteed as long as 
$m_3 \lesssim M_\bullet$ (and of course $a_3 > a$). 
As shown in Ref.~\cite{Liu2015}, the maximum orbital eccentricity that could be excited 
turns out to be $e_{\rm max} \approx \sqrt{P_{\rm pre}/P_{\rm K-L}} \ll 1$ assuming an initial circular orbit
and a stellar mass third object $m_3 \lesssim 10^2 M_\odot$.
Therefore Kozai-Lidov oscillations are not expected to drive the SMO off the low-eccentricity orbit.

\begin{figure}
\includegraphics[scale=0.11]{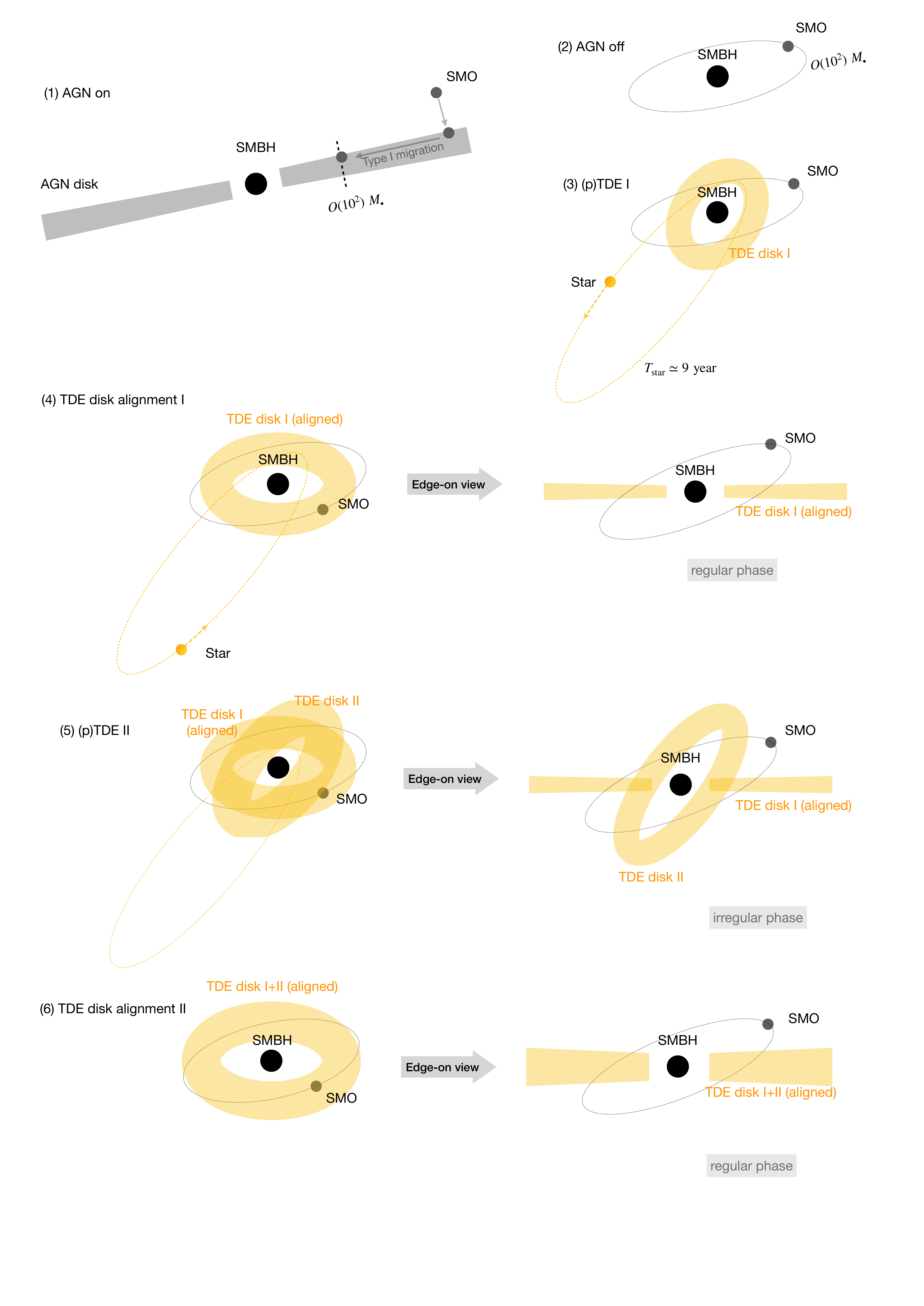}
\caption{\label{fig:cartoon} 
A cartoon plot showing the whole history of GSN 069 EMRI formation and QPE production:
(1) in a previous AGN phase, a SMO is captured on to the AGN disk
and migrates inward driven by density waves until $r=O(10^2) M_\bullet$ 
where GW radiation takes over; (2) after the AGN turns off, the SMO resides on a nearly 
circular orbit with radius $r=O(10^2) M_\bullet$; (3) a star is partially tidally disrupted and the stripped gas forms a TDE disk later; (4) the regular phase of QPEs starts when the gas settles down into a TDE disk which takes a few years to be aligned on the equator; (5) the regular QPE phase is disturbed by the second tidal disruption of the star; (6) a new regular QPE phase starts after the gas from
the second (p)TDE settles down.}
\end{figure}

To summarize, the low-eccentricity EMRI with semi-major axis $a=\mathcal{O}(10^2) M_\bullet$ (Eq.~\ref{eq:obt_pars}) in GSN 069 seems unlikely to be from the loss-cone channel or the Hills mechanism, and is consistent with the wet EMRI channel prediction.
In Fig.~\ref{fig:cartoon}, we tentatively display the whole history of GSN 069 EMRI formation in a previous AGN phase and the on-going QPEs from EMRI and TDE disk collisions.

\subsection{sBH EMRIs versus stellar EMRIs}

In the EMRI+TDE disk model, the secondary object could be a normal star \cite{Linial2023} or a sBH \cite{Franchini2023},
which predict similar QPE properties and are not easy to be distinguished from current QPE observations 
(see the above two references for more detailed arguments). 
Identifying the nature of the secondary object will be invaluable for 
accurately predicting the rate of EMRIs detectable by spaceborne GW detectors 
and for distinguishing different accretion disk models
in light of the increasing number of QPE detections.

Both models predict a decay in the orbital period due to energy loss in the collisions with different rates $\dot T_{\rm obt}$ (see Eqs.~(\ref{eq:Tdot},\ref{eq:Tdot_star})). The current constraint is not sufficiently accurate for distinguishing the two (see Eq.~(\ref{eq:Tsumdot})).
Longer monitoring the existing QPE sources is necessary in pinning down the orbital period decay rate $\dot T_{\rm obt}$,
and consequently identifying the nature of the SMO.

In the two models, the lifetime of QPE activities might be different.
In general, three timescales are relevant to the QPE lifetime \cite{Linial2023,Franchini2023}: the SMO orbital decay timescale $t_{\rm decay}$,
the TDE disk lifetime $t_{\rm TDE \ disk}$, and the star survival timescale $t_{\star, \rm sur}$ (for stellar EMRIs only).
The decay timescale $t_{\rm decay}:=|T_{\rm obt}/\dot T_{\rm obt}| = \mathcal{O}(10^{3}-10^{4})$ yr 
[Eqs.~(\ref{eq:Tdot},\ref{eq:Tdot_star})]
is much longer than  $t_{\rm TDE \ disk}=\mathcal{O}(1-10)$ yr.
The star survival timescale as interacting with the TDE disk is rather uncertain.
In Ref.~\cite{Linial2023}, $t_{\star, \rm sur}$ is estimated as the star mass ablation timescale 
\be 
t_{\star, \rm sur} :=\frac{m_\star}{\Delta m_\star}\frac{T_{\rm obt}}{2} \approx  160\  \alpha_{\rm 0.01} M_{\bullet,6}^{4/3} \dot M_{\bullet,0.1}^{2} T_{\rm obt, 20}^{2/3} m_{\star,\odot}^2 R_{\star,\odot}^{-4}  \ {\rm yr}\ ,
\ee 
within which the star loses most of its mass as it crosses the TDE disk and a small amount of gas $\Delta m_\star$ gets stripped each time due to the ram pressure. 
In this estimate, the star survives sufficiently long and the lifetime of QPE activities will be roughly $t_{\rm TDE, disk}$.
In Ref.~\cite{Franchini2023},  $t_{\star, \rm sur}$
is estimated by comparing the impact energy $\delta E_\star$ with the star binding energy $E_{\rm bind}:=Gm_\star^2/2R_\star$, 
\be 
\begin{aligned}
    t_{\star,\rm sur} &:=\frac{E_{\rm bind}}{\delta E_\star}\frac{T_{\rm obt}}{2} \\ 
    &\approx 
12\ \alpha_{0.01} \dot M_{\bullet, 0.1} M_{\bullet,6}^{1/3} T_{\rm obt,20}^{2/3} (\sin\iota_{\rm sd})^{-1} m_{\star,\odot}^2
R_{\star,\odot}^{-3}\ {\rm d}\ .
\end{aligned}
\ee 
This estimate gives a much shorter star survival time and disfavors the stellar EMRI model.
Therefore the lifetime of QPE activity is roughly equal to the TDE disk lifetime $\mathcal{O}(1-10)$ yr in the sBH EMRI model,
while is much more uncertain in the stellar EMRI model due to the uncertain star survival time. Detailed simulations of star-disk collisions are necessary for figuring out the 
star survival time. Once confirmed, the lifetime of QPE activities will be useful in distinguishing the two models.

Both models predict the strong-weak pattern in the QPE intensities, though with different dependence on the orbital eccentricity:
$\delta E_{\rm sBH} \propto \Sigma r \propto r^{5/2}$ and $\delta E_{\star} \propto \Sigma r^{-1} \propto r^{1/2}$ 
from Eqs.~(\ref{eq:delta_E_star},\ref{eq:delta_E}), therefore the strong-weak QPE intensity contrasts are $(I_{\rm strong}/I_{\rm weak}-1) \leq (r_{\rm a}/r_{\rm p}-1)^{5/2 \ {\rm or}\ 1/2} \approx 5e$ or $e$ in these two models, respectively. 
For GSN 069, the intensity contrasts are roughly $30\%$ (see Fig.~\ref{fig:raw_lc}), 
which requires an orbital eccentricity $e\approx0.06$ for the sBH EMRI and 
$e\approx0.3$ for the stellar EMRI. The EMRI orbital eccentricity obtained from the QPE timing is $e=0.04^{+0.04}_{-0.03}$
at 2-$\sigma$ confidence level, which favors the sBH EMRI. But this inference depends on the standard $\alpha$ disk assumption,
with the disk surface density $\Sigma \propto r^{3/2}$, the accuracy of which is not yet confirmed for TDE disks.
For example, in the $\beta$ disk model where $\Sigma \propto r^{-8/5}$ \cite{Kocsis2011}, the conclusion will be opposite.
An important consequence of this dependence on the disk surface density profile is that the QPEs can be used as a probe to different accretion disk models as long as the 
nature of the SMO is confirmed, say, via the orbital change rate $\dot T_{\rm obt}$ or the star survival time $t_{\star,\rm sur}$ as explained in the previous paragraphs.

In the above intensity analysis, we have implicitly assumed the symmetry of the emissions about the mid-plane of the disk after being shocked by the SMO from either direction, i.e., $I_{\downarrow v}^{\uparrow \gamma}=I_{\downarrow v}^{\downarrow \gamma}$
and $I_{\uparrow v}^{\uparrow \gamma}=I_{\uparrow v}^{\downarrow \gamma}$, where $\updownarrow v$ denote the SMO moving directions and the $\updownarrow \gamma$ denote the emission directions.
The (approximate) symmetry has been verified in local simulations of sBH-disk collision \cite{Ivanov1998}, but may not be true for star-disk collisions, where 
the emission could be preferably on one side of the post-collision disk due to the large geometrical size of the star. 
Such kind of asymmetry is indeed observed in our global HD simulations with different geometrical sizes of SMO colliding with the disk (see Appendix~\ref{app:hd_sim} for more details). Specially, with a large softening/sink particle radius to mimic the star-disk collision, the integrated density/temperature perturbations from the lower and upper disks behaves more asymmetric than the smaller softening case. But this kind of asymmetric is not as that strong as we expect.  We suspect that this could be due to that we haven't explore the extremely large geometry size contrast for these two scenarios, which is unlikely feasible in our global HD simulations.

In the above intensity analysis, we have also assumed a steady axisymmetric accretion disk. In fact, the TDE disk
might be lopsided or eccentric \cite{Dai2021}. Though the non-axisymmetric nature of the disk does not change the alternating strong-weak pattern in the QPE intensities, it is expected to modulate 
the QPE intensities on the SMO orbital precession timescale as the collision positions vary.

\subsection{Model imitations and future work}
In this work, we have been focusing on the analysis of the QPE source GSN 069. 
In principle, one can conduct a full parameter inference on the EMRI orbital model and the flare emission model.
In fact, it is not straightforward to accurately model the QPE light curves with simple emission models. 
In this work, we used an expanding plasma ball model and a phenomenological model, both of which might be subject to some 
systematics in determining the flare starting times $t_0^{(k)}\pm \sigma(t_0^{(k)})$
(see discussion about identifying possible systematics from HD simulations in Appendix~\ref{app:hd_sim}). 
To mitigate the impact of these potential systematics,  we multiple the uncertainties by a scale factor $F_t$ in inferring the EMRI orbital parameters from the flare starting times.  Another degree of freedom we did not consider in this work is the SMBH spin, which drives Lense-Thirring precession and consequently modulates the QPE recurrence times. 

We will improve these model limitations and apply the improved analysis on all the existing QPE sources in a follow-up work, where the SMBH spin is straightforward to take into consideration 
in the EMRI orbits, and the emission model systematics may be improved with a simulation motivated model and/or a
hierarchical inference method \cite{Isi2019}  widely used in the GW community.

Recently, QPEs with recurrence times of $\sim $ 3 weeks was found in source Swift J0230+28 \cite{Evans2023,Guolo2024},
and it is interesting to see whether these QPEs fit in the same framework.

\vspace{0.2 cm}

\acknowledgments
We thank the referee for carefully reading this paper and providing encouraging and valuable comments.
We thank  Liang Dai, Dong Lai and Bin Liu for enlightening discussions. We also thank Jialai Kang, Shifeng Huang, Giovanni Miniutti for helpful discussions on the X-ray data analysis and Qian Hu for valuable discussion. Lei Huang thanks for the support by National Natural Science Foundation of China (11933007,12325302), Key Research Program of Frontier Sciences, CAS (ZDBS-LY-SLH011),  Shanghai Pilot Program for Basic Research-Chinese Academy of Science, Shanghai Branch (JCYJ-SHFY-2021-013).
Y.P.L. is supported in part by the Natural Science Foundation of China (grants 12373070, and 12192223), the Natural Science Foundation of Shanghai (grant NO. 23ZR1473700). The calculations have made use of the High Performance Computing
Resource in the Core Facility for Advanced Research Computing
at Shanghai Astronomical Observatory.

This paper made use of data from XMM-Newton, an ESA science mission with instruments and contributions directly funded
by the ESA Member States and NASA. 
\appendix

\section{Constraints on the parameters of the emission model and the flare timing model}\label{app:corner}

\begin{figure*}
\includegraphics[scale=0.30]{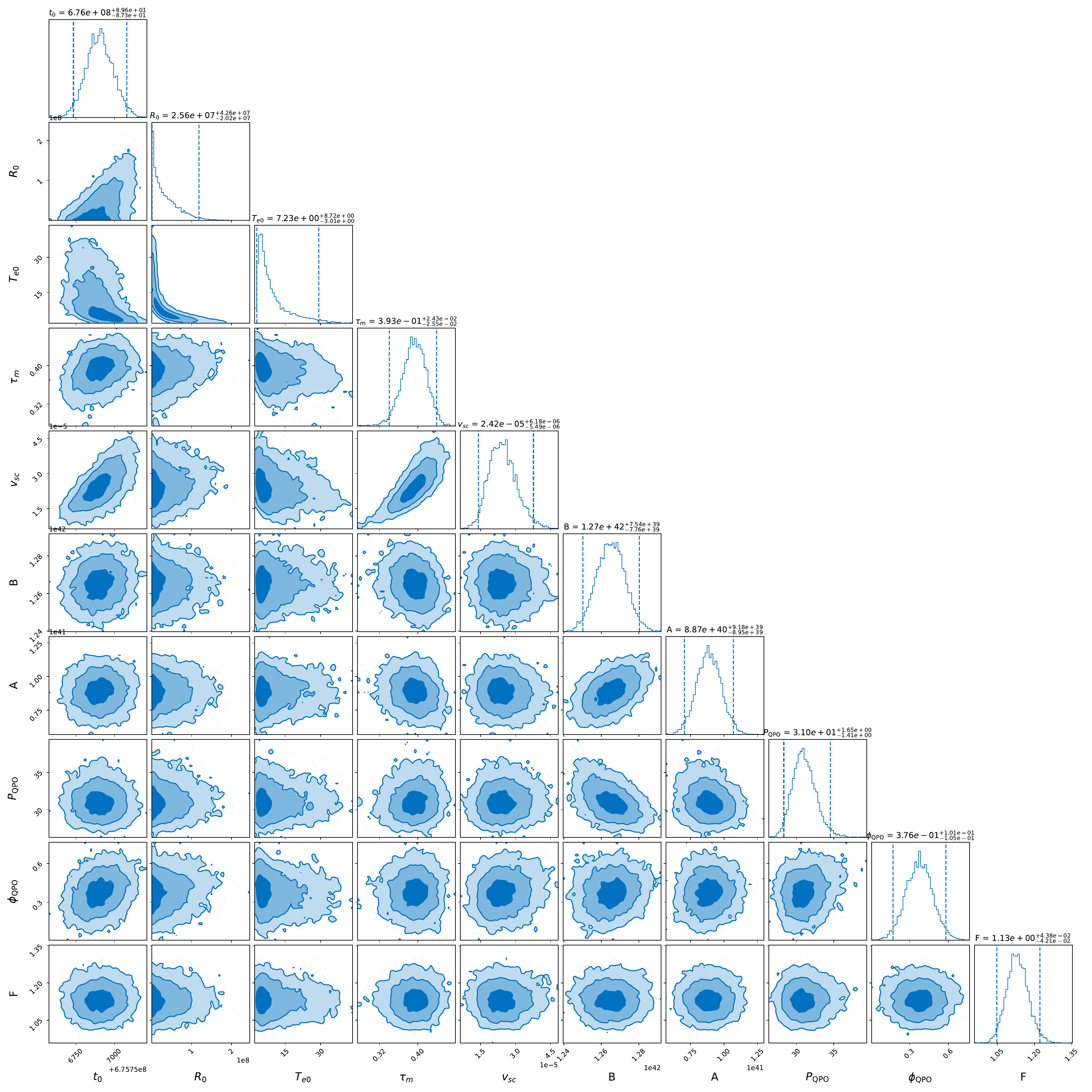}
\caption{\label{fig:eg_corner} The corner plot of the emission model parameters of the second flare in XMM5: $t_0 [{\rm sec}], R_0 [{\rm cm}], T_{e0} [{\rm keV}], \tau_m [{\rm hour}], v_{sc} [{\rm c}], B [{\rm erg/s}], A [{\rm erg/s}], P_{\rm QPO} [{\rm ks}], \phi_{\rm QPO}, F$, where each pair of vertical lines denotes the 2-$\sigma$ confidence level.}
\end{figure*}

\begin{figure}
\includegraphics[scale=0.45]{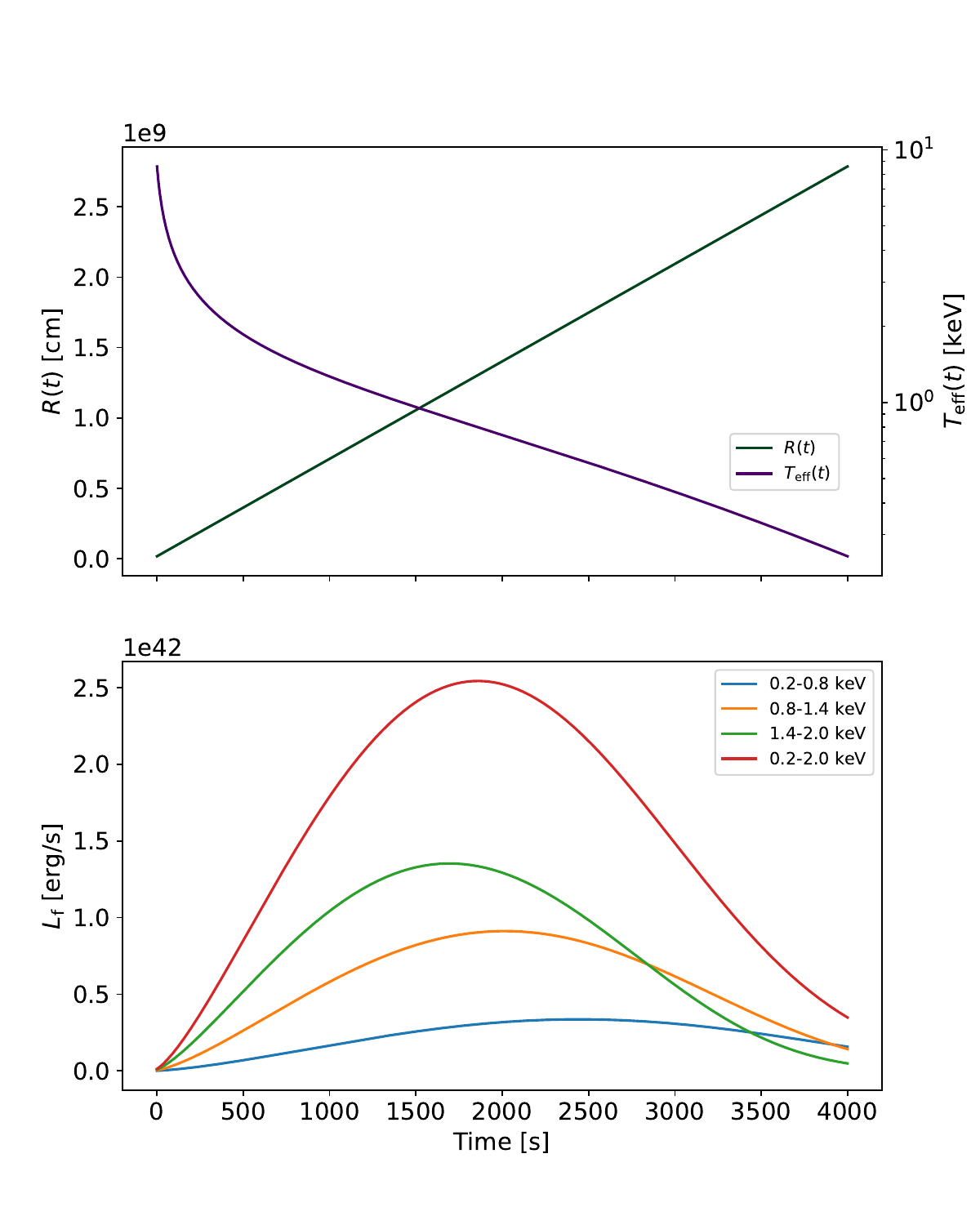}
\caption{\label{fig:evolution} Details of the best fit model of the second flare in XMM5.
Top panel: the evolution of plasma ball radius and effective temperature $R(t), T_{\rm eff}(t)$; 
bottom panel: spectral evolution of the plasma ball emission, 
where eruptions measured in higher energy bands are stronger, peak earlier, and have shorter
duration than when measured at lower energies}
\end{figure}

Fig.~\ref{fig:eg_corner} displays the posterior corner plot of the plasma ball model parameters for the second flare in XMM 5. 
In Fig.~\ref{fig:evolution}, we show the evolution of the plasma ball size $R(t)$
and the effective temperature $T_{\rm eff}(t)$, and its spectral evolution
in the best-fit model of the second flare in XMM5. 
Around the peak of the QPE luminosity, the plasma ball  effective temperature is $\approx 0.5$ keV, 
which is about  higher than the observed value by a factor of $2\sim 3$, though the uncertainty in the initial temperature is large
(see Fig.~\ref{fig:eg_corner}). This tension implies the limitation of the simple 
expanding plasma ball emission model. This is one of the reasons we consider an alternative phenomenological model 
for fitting the QPE light curves.

\begin{figure*}
\includegraphics[scale=0.30]{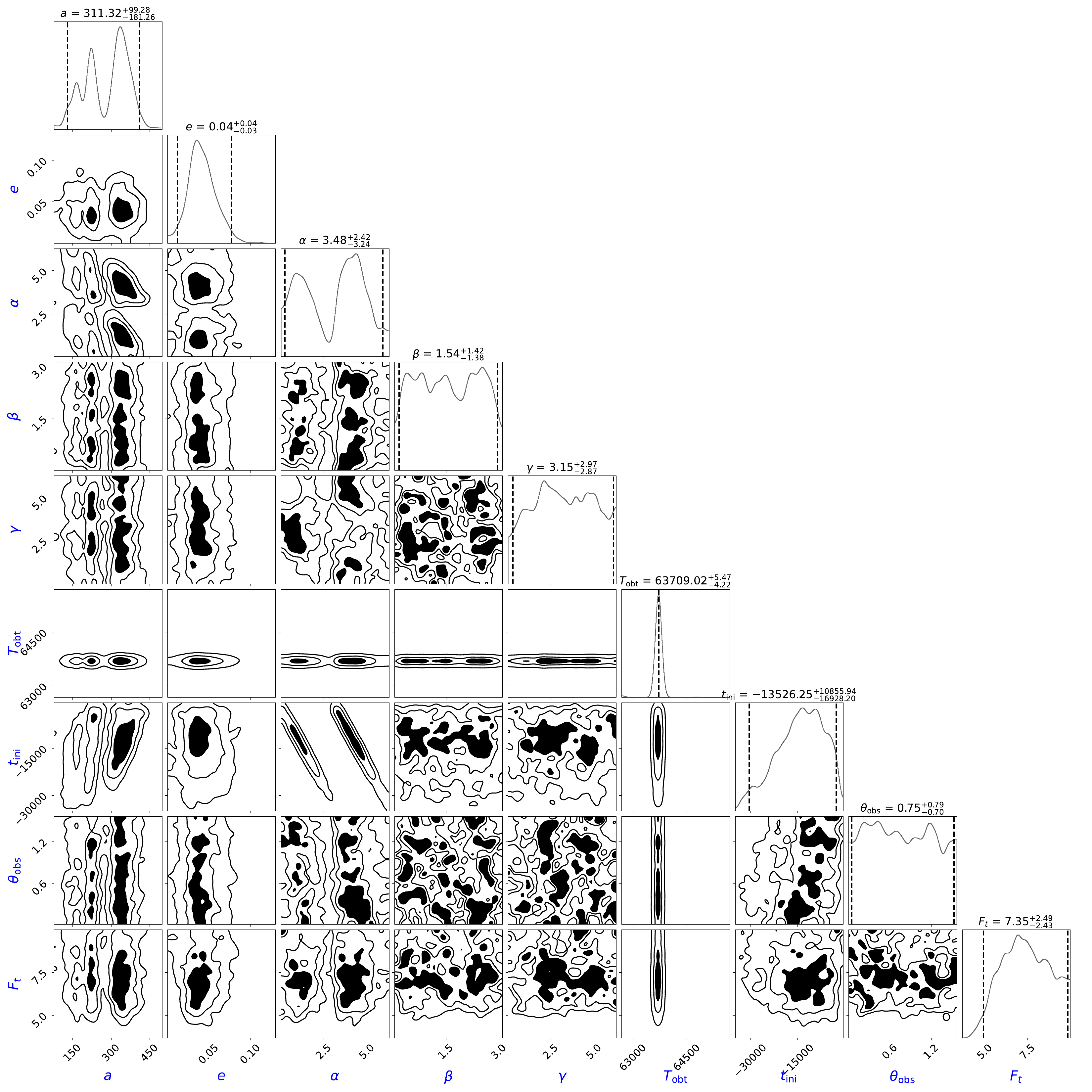}
\caption{\label{fig:obt_corner} The posterior corner plot of the flare timing model parameters: $a [M_\bullet], e, \alpha, \beta, \gamma, T_{\rm obt} [{\rm sec}], t_{\rm ini} [{\rm sec}], \theta_{\rm obs}, F_t$, where each pair of vertical lines denotes the 2-$\sigma$ confidence level. And the data used are the flare starting times $t_0^{(k)}$ shown in Table~\ref{tab:t_0_P_QPO}.} 
\end{figure*}

Fig.~\ref{fig:obt_corner} displays the  posterior corner plot of the EMRI orbital parameter constrained by the flare 
starting times $t_0^{(k)}$ shown in Table~\ref{tab:t_0_P_QPO}. All the angles defining the orbital plane orientation and the 
los direction are not well constrained saturating their priors, but the posteriors of the intrinsic orbital parameters yield
import clues of the EMRI formation history as explained the Summary section.

\begin{figure*}
\includegraphics[scale=0.30]{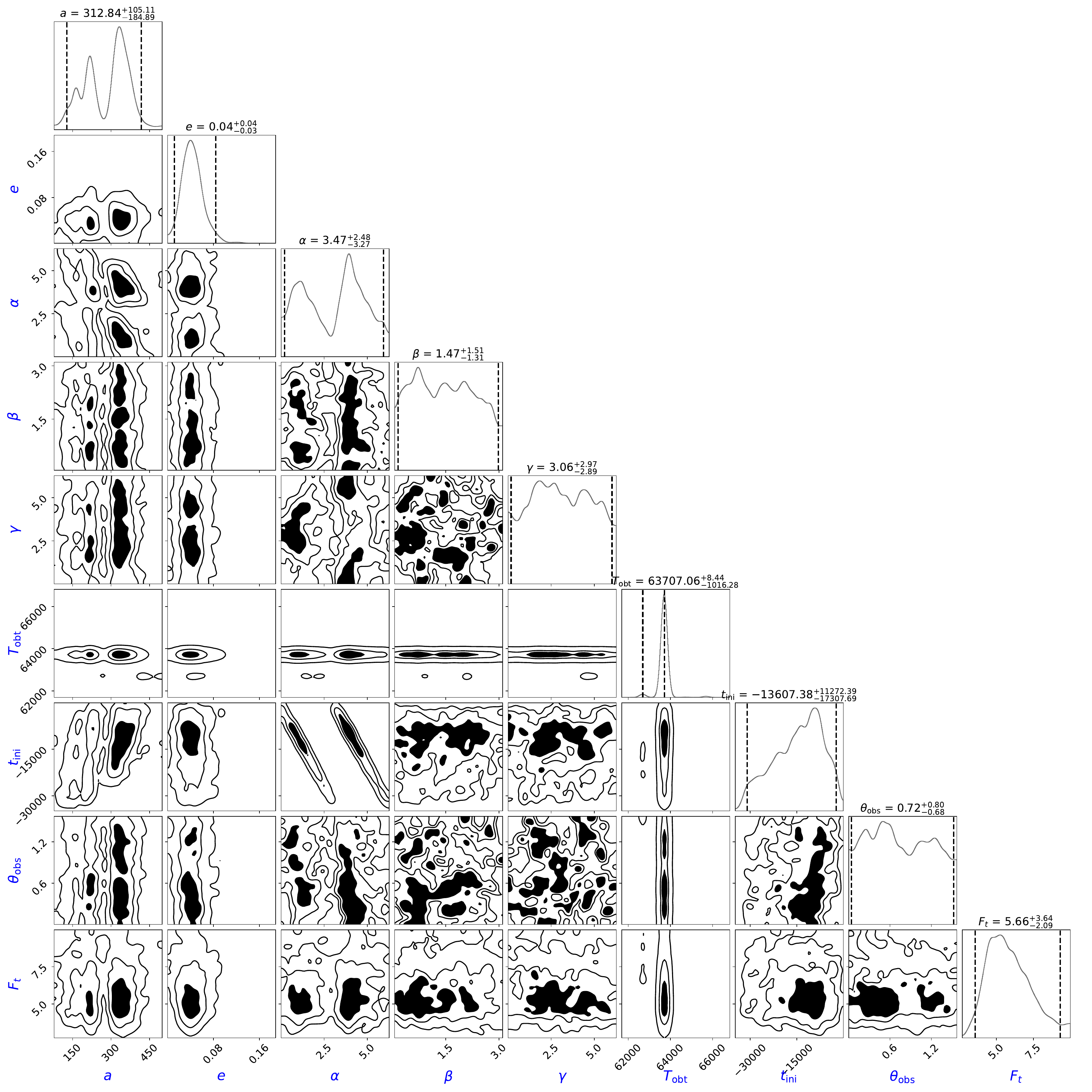}
\caption{\label{fig:obt_corner_phen} Similar to Fig.~\ref{fig:obt_corner}, except the data used are 
the flare starting times $t_0^{(k)}$ in Table~\ref{tab:t_0_P_QPO_phenom}.} 
\end{figure*}

\section{Scatter experiments of the remnant star and the SMO}\label{app:scatter}

If the new period $T^{\rm (new)}_{\rm sum}$ in XMM 12 turns out to be largely different from those in the old regular phase
and in the irregular phase,
this new phase will pose a challenge to many existing models, including the EMRI+TDE disk model.
There was some speculation that the shorter $T^{\rm (new)}_{\rm short}$ and $T^{\rm (new)}_{\rm sum}$ are the result
of a large change in the EMRI orbital ($\sim 10\%$ decrease in the semi-major axis and $\sim 0.2$ increase in the eccentricity) after a close encounter with the remnant star at its pericenter during TDE 2.
We examine this speculation by scattering experiments, and we find such large orbital change seems unlikely.


In order to test whether the orbital eccentricity of the SMO can be excited during the close encounter with the remnant star, we performed scatter experiments with \textit{N}-body simulations to follow the orbital evolution of the sBH or the star during and after the TDE 2 event. 

We use a 4th order Hermite integrator with block time step \citep{Kokubo_1998} to calculate the orbital evolution of the two bodies. 
We first consider the case where a remnant star that has experienced TDE 1 and a sBH are orbiting a SMBH.
The mass of the remnant star, the sBH, and the SMBH are set to $2.5~M_\odot$, $10~M_\odot$, and $10^6~M_\odot$ respectively. 
The remnant star has a pericenter distance of $100~M_\bullet$ and an orbital period of 9 years, which yields an eccentricity of 0.9954. The orbital inclination of the star is set to 0.
The sBH has an initially low eccentricity of 0.05 and a semi-major axis of $160~M_\bullet$. The orbital inclination of the sBH is randomly chosen from 0 to $\pi$. 
The other orbital elements (the longitude of ascending node, the argument of pericenter, and the time of pericenter passage) are randomly chosen from 0 to $2\pi$.
We also consider the case where the remnant star encounters with another star instead of a sBH near the pericenter. 
In this case, the sBH is replaced by a star of $1~M_\odot$. The orbital parameters remain unchanged.

In both cases, we integrate the system for roughly half an orbit of the remnant star to capture the orbital change of the SMO before and after the close encounter with the remnant star at the pericenter.
A softening parameter $\epsilon \simeq 2~R_\odot$ is used for close encounter. 
1000 simulations are performed in each case. No close encounters lead to a $10\%$ decrease in the orbital semi-major axis
or a $0.2$ increase in the orbital eccentricity.

\section{HD simulations of SMO-disk collisions} \label{app:hd_sim}

We carry out a few 3D hydrodynamical simulations for the SMO-disk collision using Athena++ \citep{Stone2020}. The thin disk is initialized with an aspect ratio of $h/r=0.03$, and an $\alpha$-viscosity \citep{SS1973} is implement with $\alpha=0.1$.
For simplicity, the SMO collides with the disk around the central SMBH vertically. The SMO is initially far above the midplane such that the gravitational interaction between the SMO and the disk is weak. The motion of the SMO is prescribed by only vertical velocity $-v_{\rm k}$ while fixing cylindrical $R=r_{0}$ and azimuth $\phi=0$ location in time, where $v_{\rm k}$ is the local Keplerian velocity. 

The gravitational potential of the SMO is softened with the classical Plume potential with a softening scale of $\epsilon$ and the accretion of the SMO is modelled as a sink particle with the same softening radius $\epsilon$.  The softening/sink radius is to mimic the physical size of SMO.  
We adopt different softening/sink radii ($\epsilon=0.1\ R_{\rm H}$ and $\epsilon=0.3\ R_{\rm H}$) to quantify the effect of the different physical size of SMO on the collision-induced emission, where $R_{\rm H}$ is the Hill radius of the colliding object. With a mass ratio of $q=10^{-3}$, $R_{\rm H}=0.07\ r_{0}$, this leads to $\epsilon=0.007\ r_{0}$ and $\epsilon=0.02\ r_{0}$, respectively. A relative large mass SMO is adopted to save the computational cost as it is very challenging to well resolve the sink radius of the realistic mass ratio object even with grid refinement, e.g., $q\lesssim 10^{-5}$. The softening scale adopted here is still too large compared to the size the sBH, which is extremely too small to simulate numerically. 
For stellar-mass black hole collision, the sink hole radius could be as small as the event horizon of the sBH, and orders of magnitude smaller than the Bondi/Hill radius of sBH.
The case with a larger softening size is adopted for the case of star-disk collision, for which the physical size of star is usually much larger than the Bondi radius.

We evolve the gas adiabatically with an adiabatic index $\gamma=4/3$. The disk is resolved with a root grid of $[n_{r}, n_{\theta}, n_{\phi}]=[128,16,512]$, where the radial domain is $[0.5,2.5]r_{0}$, and 3.5 disk scale height is modelled in the $\theta$ direction. Three levels of static refinement with a refine sphere of $\delta r=0.07\ r_{0}$ around the midplane is adopted to well resolve the collision location. As such, we can well resolve the Hill radius of the SMO with 40 grids in each dimension.

\begin{figure*}
\includegraphics[clip=true, width=0.9\textwidth]{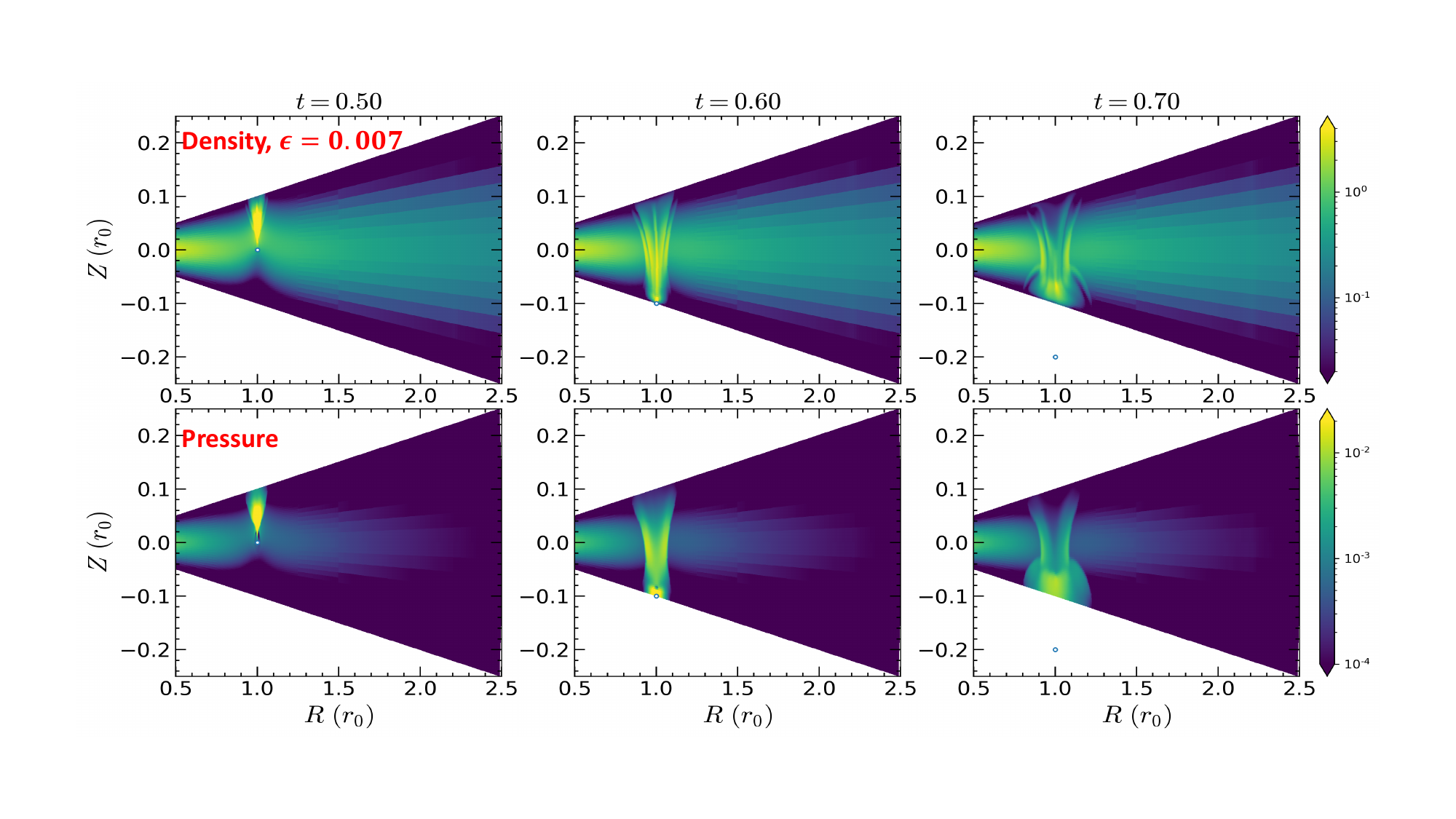}
\caption{Density (upper panels) and pressure (lower panels) distribution in the $R-Z$ plane with a slice along the azimuth of the colliding object, i.e., $\phi=0$. The colliding object has a softening/sink radius of $\epsilon=0.007\ r_{0}$. Different columns correspond to different times in unit of the local Keplerian orbital period. The small open circle denotes the position of the colliding object.} \label{fig:sigma_sft007}
\end{figure*}

\begin{figure*}
\includegraphics[clip=true, width=0.9\textwidth]{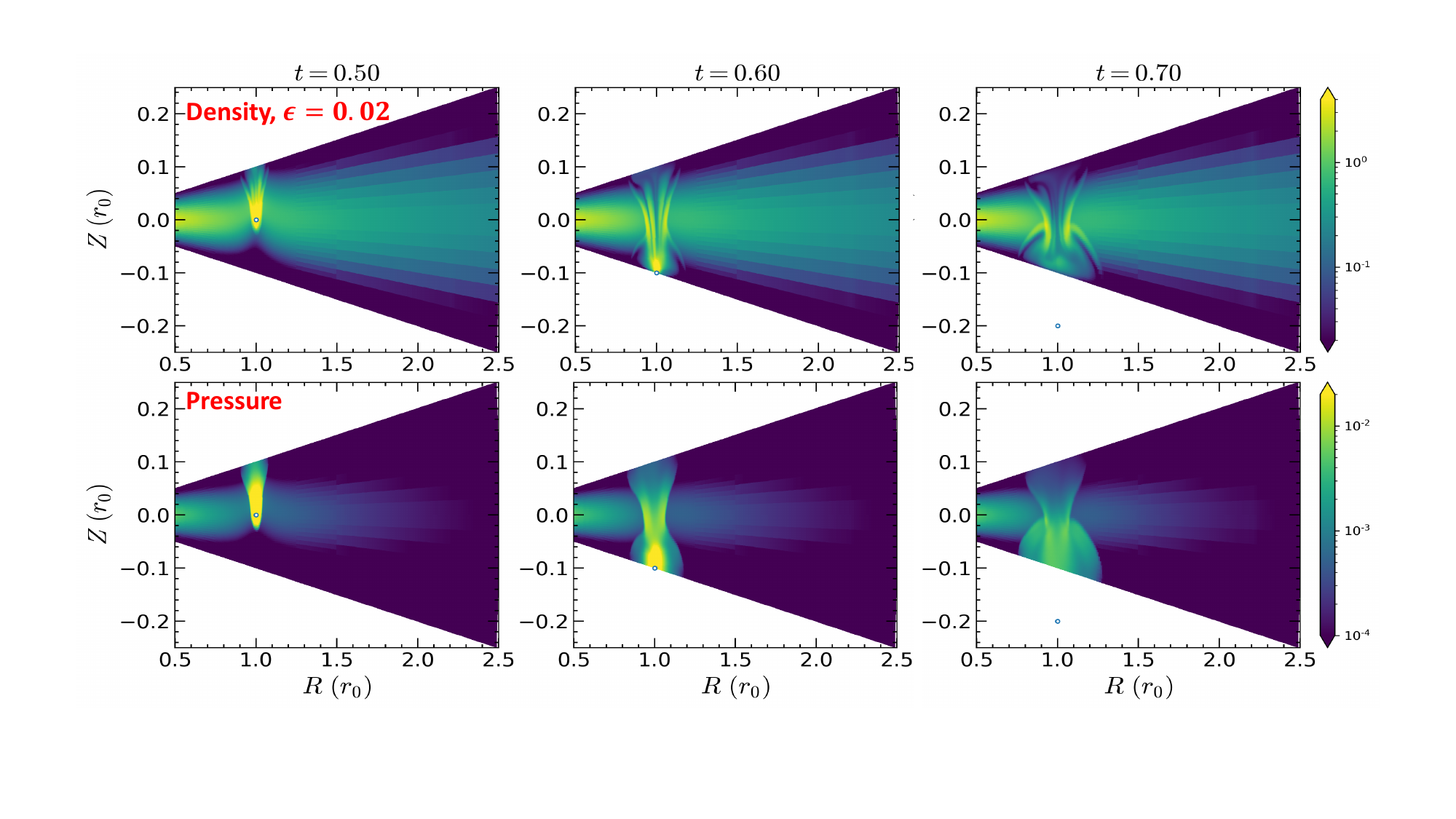}
\caption{Similar to Fig.~\ref{fig:sigma_sft007}, but with $\epsilon=0.02\ r_{0}$} \label{fig:sigma_sft02}
\end{figure*}

After the SMO-disk collision, there exists strong shocks which heat the gas around a narrow band of the collision site. By checking the vertical density and pressure distribution as shown in Fig.~\ref{fig:sigma_sft007} and Fig.~\ref{fig:sigma_sft02}, the post-collision perturbation is asymmetric above and below the disk midplane. There is a dense blob  above the midplanet before $t=0.5$, i.e., prior to the collision at the midplane. The hot blob expands radially and vertically suffering from shearing motion of the disk, which can induce spiral arms in the disk. The perturbation then becomes stronger at the lower half plane of the disk, i.e., after the collision at the midplane.   This asymmetric pattern is slightly stronger for the star-disk collision with a larger $\epsilon$ (at time $t=0.6$ of Fig.~\ref{fig:sigma_sft007} and Fig.~\ref{fig:sigma_sft02}).

\begin{figure}
\includegraphics[clip=true, width=0.45\textwidth]{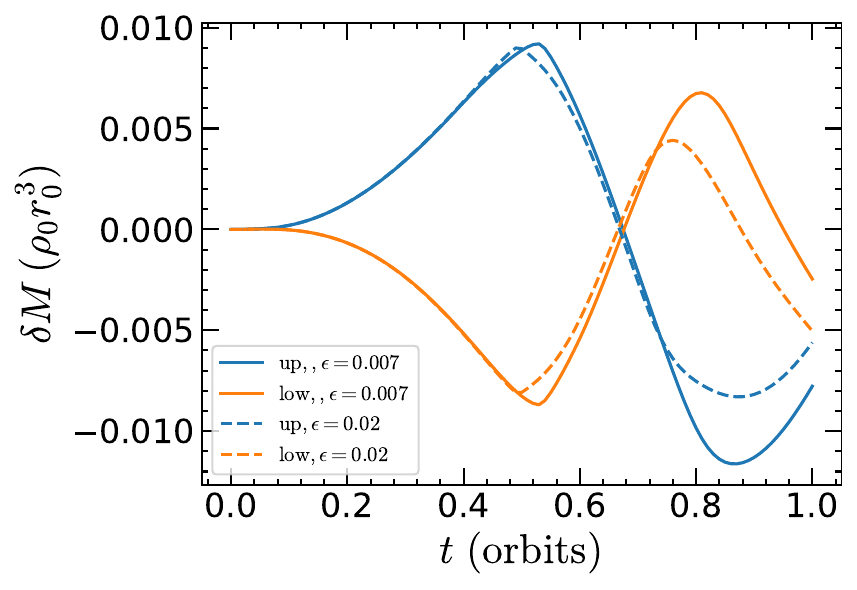}
\caption{Perturbed disk mass (subtracted from the initial disk mass) as a function of time. Solid lines correspond to the case with a softening/sink radius of $\epsilon=0.007\ r_{0}$, while the dashed lines are the case of $\epsilon=0.02\ r_{0}$. Blue (orange) lines are integrated over the upper (lower) half of disk.} \label{fig:dm_sim}
\end{figure}

The shocked-heated dense blob will be responsible for the X-ray emission in observations. To quantify the emission features, we calculate the perturbed disk mass around the collision site $r=[0.8,1.2]\ r_{0}$, we integrate the disk mass from the upper and lower half of the disk for two different softening/sink radius, shown in Fig.~\ref{fig:dm_sim}. It is clearly seen that the perturbation is most prominent in the upper half of the disk before the collision and then becomes stronger later at the lower half of the disk. As expected, there is a time-delay between the peak of perturbed disk mass for the upper and lower disk, though this time delay may suffer from the boundary condition effect in the $\theta$ direction as we do not simulate the full $\theta$ domain of the disk. The difference of the integrated mass is not that large,  although the local asymmetry of the density/pressure shown in Fig.~\ref{fig:sigma_sft007} and Fig.~\ref{fig:sigma_sft02} is stronger.

In addition, along the moving direction of the SMO, the burst actually consists of two components:
a ``precursor'' burst at the moment of the shock breaking out the disk surface ($t=0.6$ in Fig.~\ref{fig:sigma_sft007} and Fig.~\ref{fig:sigma_sft02}) followed by a main burst sourced by the heated gas in the shocked column. 
In either light curve model, the former component is not modelled and is a possible reason of $F_t > 1$.

\bibliography{ms}
\end{document}